\documentclass[aps,pra,nofootinbib,superscriptaddress,twocolumn]{revtex4-1}

\usepackage[utf8,latin1]{inputenc}
\usepackage[T1]{fontenc}     
\usepackage[british]{babel}
\usepackage[dvipsnames]{xcolor}
\usepackage{graphicx}
\usepackage{epsfig}
\usepackage{dsfont}
\usepackage{lipsum}     
\usepackage{physics}     
\usepackage{algorithm2e}
\usepackage{color}
\usepackage{lmodern}
\usepackage[babel=true]{microtype}
\usepackage{amsmath,amssymb,amsthm}
\usepackage{mathtools}
\usepackage[caption=false]{subfig}
\usepackage{bm}
\usepackage{soul}
\usepackage[toc,page]{appendix}
\usepackage{enumerate}
\usepackage[colorlinks=true,citecolor=Green,linkcolor=Red,urlcolor=Cyan,hyperindex]{hyperref}
\usepackage{cleveref}
\usepackage{hyperref}

\newcommand{\cket}[1]{\left\vert#1\right)}
\newcommand{\cbra}[1]{\left(#1\right \vert}
\newcommand{\cbraket}[2]{\left(#1\right \vert\left. #2\right)}

\newcommand{\cL}{\mathcal{L}}
\newcommand{\sgn}{\text{sgn}}
\newcommand{\rulesep}{\unskip\ \vrule\ }
\theoremstyle{definition}
\newtheorem{definition}{Definition}[]
\allowdisplaybreaks

\newtheorem{theorem}{Theorem}
\begin{document}
  
  \title{Circuit connectivity boosts by quantum-classical-quantum interfaces}
\author{Roeland Wiersema}
\email{rwiersema@uwaterloo.ca}
\affiliation{Vector Institute, MaRS  Centre,  Toronto,  Ontario,  M5G  1M1,  Canada}
\affiliation{Department of Physics and Astronomy, University of Waterloo, Ontario, N2L 3G1, Canada}

\author{Leonardo Guerini}
\affiliation{Department of Mathematics, Federal University of Santa Maria, Santa Maria, RS 97105-900, Brazil}
\affiliation{Instituto de F\'isica, Federal University of Rio de Janeiro, P. O. Box 68528, Rio de Janeiro 21941-972, Brazil}

\author{Juan Felipe Carrasquilla}
\affiliation{Vector Institute, MaRS  Centre,  Toronto,  Ontario,  M5G  1M1,  Canada}
\affiliation{Department of Physics and Astronomy, University of Waterloo, Ontario, N2L 3G1, Canada}
\affiliation{ Department of Physics, University of Toronto, Ontario M5S 1A7, Canada}

\author{Leandro Aolita}
\affiliation{Instituto de F\'isica, Federal University of Rio de Janeiro, P. O. Box 68528, Rio de Janeiro 21941-972, Brazil}
\affiliation{Quantum Research Centre, Technology Innovation Institute, Abu Dhabi, United Arab Emirates}

\date{\today}

\begin{abstract}
High-connectivity circuits are a major roadblock for current quantum hardware. We propose a hybrid classical-quantum algorithm to simulate such circuits without swap-gate ladders. As main technical tool, we introduce quantum-classical-quantum interfaces. These replace an experimentally problematic gate (e.g. a long-range one) by single-qubit random measurements followed by state-preparations sampled according to a classical quasi-probability simulation of the noiseless gate. Each interface introduces a multiplicative statistical overhead which is remarkably independent of the on-chip qubit distance. Hence, by applying interfaces to the longest range gates in a target circuit, significant reductions in circuit depth and gate infidelity can be attained. We numerically show the efficacy of our method for a Bell-state circuit for two increasingly distant qubits and a variational ground-state solver for the transverse-field Ising model on a ring. Our findings provide a versatile toolbox for error-mitigation and circuit boosts tailored for noisy, intermediate-scale quantum computation. 
\end{abstract}
  \maketitle
  
  \section{Introduction}
 Quantum computation promises a major disruption in high-performance computing, with  applications on diverse fields ranging from many-body physics and chemistry to machine learning, finance, automation, or logistics, to name a few \cite{Q_alg_review_Montanaro,Q_Machine_Learning_rev, Q_Finance_review}.  However, the current paradigm of noisy, intermediate-scale quantum (NISQ) devices limits quantum algorithms to circuits of low qubit numbers, low depth, and low connectivity \cite{Preskill2018nisq}. This poses serious concerns on the actual usefulness of quantum computers in the near term and has thus ignited a both experimental and theoretical quest for ways to unleash the potential of quantum algorithms with NISQ hardware \cite{Par_Q_circuits19,NISQ_review,Variational_q_algs_review}.
 
A large class of NISQ algorithms are based on hybrid quantum-classical approaches. One of the most succesful of these consists of parametrized quantum circuits variationally optimized through a classical optimizer aimed at approximating a target ground state \cite{QAOA,Peruzzo2014vqe}. To combat the noise in these systems, subsequent variants incorporated the idea of quantum error mitigation \cite{Temme2017errormit, Endo2018errormit, Yi2017activeerror, Kandala2019errormit}. This refers to schemes whereby noisy experimental implementations (e.g., at different noise regimes or with different gate choices), together with suitable classical post-processing, are used to simulate a target, noiseless quantum circuit of limited size. This offers a NISQ alternative to quantum error correction (which requires large-scale quantum circuits), where full fault tolerance is achieved by actively correcting errors on the quantum hardware during the execution of the computation.  

More recently, a different type of hybrid method has been put forward  \cite{Bravyi2016trading,Dunjko18,Peng2020smallqc,Tang2021circuitcut,Perlin2021circuitcut}. There, a classical algorithm calls a quantum computer as a sub-routine to simulate a larger quantum circuit. However, the cost of this is that both the number of queries to the quantum sub-routine and the classical post-processing runtime unavoidably grow exponentially with the size of the target circuit. Moreover, a particularly challenging aspect of NISQ devices is their inability to run algorithms that require high, long-range connectivity among the constituent qubits. In most NISQ hardwares, long-range gates are synthesized by a long sequence of nearest-neighbor gates. This drastically inflates the circuit depth and causes large infidelity due to noise accumulation incurred during the syntheses. This is a crucial limitation in the NISQ era.

Here, we take a conceptually different direction from previous hybrid schemes: instead of assembling a large quantum circuit from small pieces, we simulate a high-connectivity circuit from circuits with low connectivity and depth. To that end, we introduce the notion of \emph{quantum-classical-quantum} (QCQ) interfaces. A QCQ interface for a gate $U$ corresponds to a local measurement on the qubits on which $U$ acts followed by a re-preparation of those same qubits in a random product state that depends on $U$. In other words, the interface performs a hybrid quantum-classical simulation of $U$. Each interface introduces a multiplicative statistical overhead that, as we prove below, is independent of the on-chip distance between the qubits. Hence, for fixed number of interfaces, e.g., the longer the range of the target gates is, the more drastic the reduction in depth attained is at the expense of a constant overall statistical overhead. 

More technically, our interfaces combine state-of-the-art state estimation based on single-qubit random measurements \cite{Carrasquilla2019genrecon,Huang2020shadow} with quasi-probability representations based on frames \cite{Ferrie2008quasiprob, Ferrie2011quasiprob}. Such representations have been used for classically simulating a quantum circuit with Monte Carlo sampling techniques \cite{Mari_Eisert12,Veitch_12,Pashayan2015quasiprobs}. In particular, our algorithm can be seen as a hybrid version of the scheme of Ref. \cite{Pashayan2015quasiprobs} where everything is quantum except for a subset of gates that one wishes to ``cut out'' of the experimental circuit. Here, we choose such subset in terms of the on-chip qubit distance. However, other relevant choices may be due simply to error mitigation or hardware-specific limitations.
As most quasi-probability schemes, our method suffers from the infamous sign problem \cite{Hatano1992sign,Loh1990signprob,Troyer2005signprob}. Remarkably, the severity of the problem depends only on the number of interfaces and not the on-chip distance between the qubits. Moreover, as by-product contribution, in order to minimize the average sign of our quasi-probability representation, we develop a Metropolis-Hastings simulated-annealing algorithm based on random walks in the space of dual frames. We implement such walks through a convenient, long-known parametrization of generalized inverse matrices \cite{Rao1967geninv}. This allows us to decrease the sample complexity overhead per interface by almost a factor of two relative to the canonical-frame choice, constituting a practical tool of general relevance for sign-problem mitigation \cite{Marvian2019curing,Hangleiter2020easing}.

The paper is organized as follows. In Sec. \ref{sec:prelims} we introduce our notation and the necessary concepts from frame theory to understand QCQ interfaces. We then present our algorithm in detail in Sec. \ref{sec:quasiprobs}. In Sec. \ref{sec:numexp} we perform numerical experiments to show the efficacy of our method on two illustrative circuits, namely the preparation of a Bell state between two increasingly distant qubits and a variational ground-state solver for the 1D transverse-field Ising model with periodic boundary conditions. We end with a discussion of our results in Sec. \ref{sec:final} and provide a perspective on other potential applications of our method.

\section{Preliminaries \label{sec:prelims}}
  \begin{figure*}[htb!]
    \subfloat[\label{fig:mreprep_1}]{
    \centering
    \includegraphics[width=0.2\textwidth]{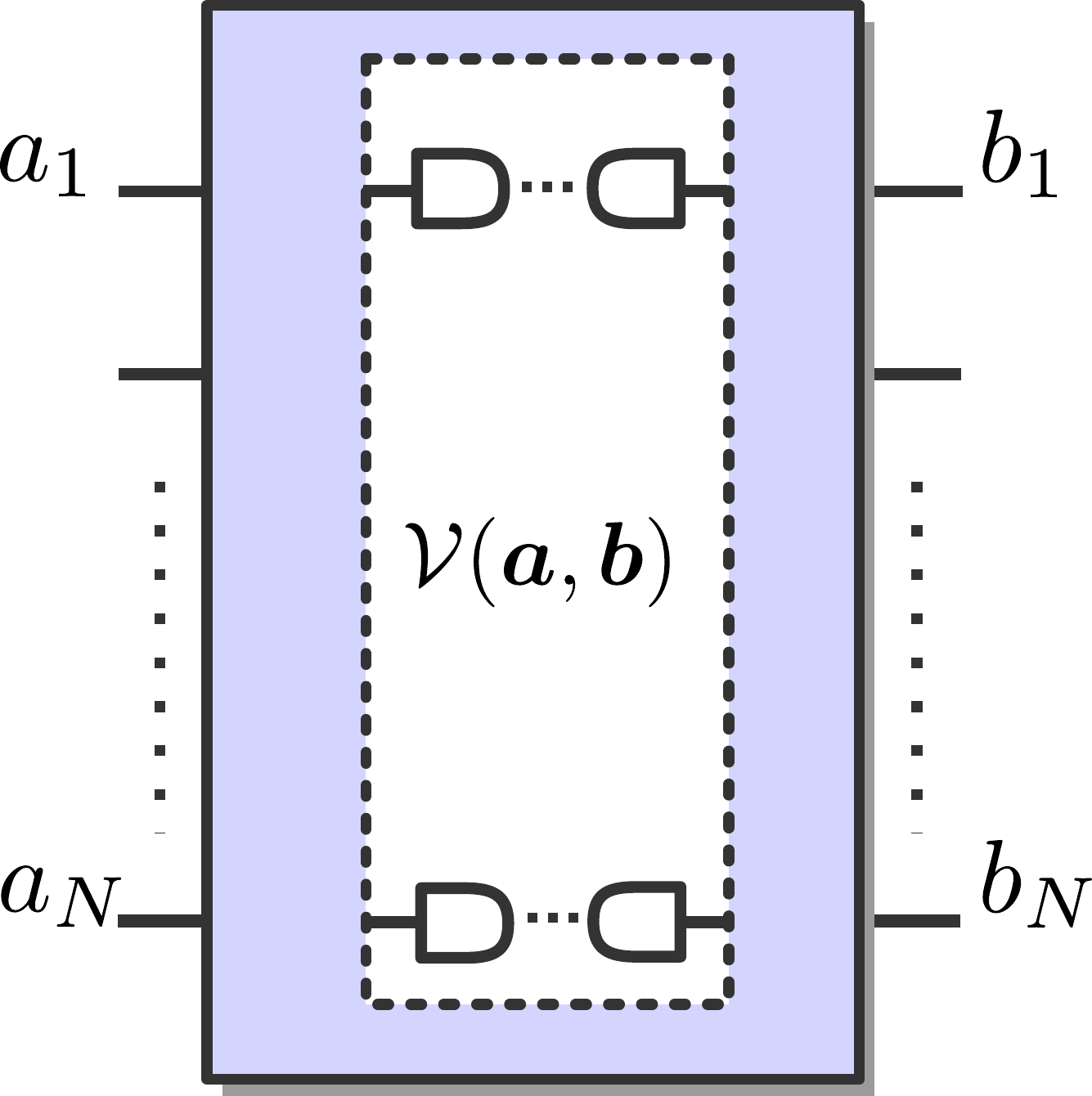}}
    \hspace{0.25cm}
    \rulesep
    \hspace{0.25cm}
    \subfloat[\label{fig:mreprep_2}]{
    \centering
    \includegraphics[width=0.7\textwidth]{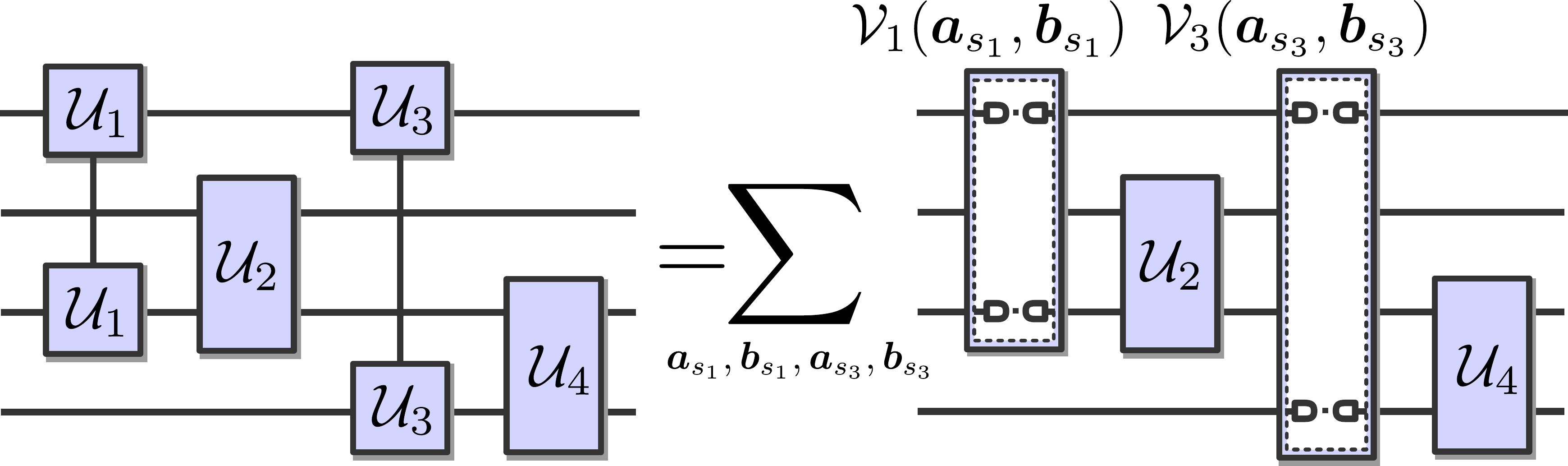}}
    
    \caption{Schematics of our method. (a) A QCQ interface $\mathcal{V}(\boldsymbol{a},\boldsymbol{b})$ simulates a gate between qubits 1 and $N$. The two qubits are measured in random single-qubit bases and reprepared in a random product state that depends on the simulated gate. The other $N-2$ qubits are left untouched. (b) An exemplary 4-qubit circuit (left) is simulated by a hybrid quantum-classical circuit (right), where the non nearest-neighbor gates $\mathcal{U}_1$ and $\mathcal{U}_3$ are substituted by QCQ interfaces [$\mathcal{V}_1(\boldsymbol{a}_{s_1},\boldsymbol{b}_{s_1})$ and $\mathcal{V}_3(\boldsymbol{a}_{s_3},\boldsymbol{b}_{s_3})$, respectively]. The summation over $(\boldsymbol{a}_{s_1},\boldsymbol{b}_{s_1}, \boldsymbol{a}_{s_3},\boldsymbol{b}_{s_3})$ represents the average over all interface outcomes sampled (see text).
}
    
    \label{fig:unitary}
  \end{figure*}
 We consider an $N$-qubit system $\mathcal{S}$ of Hilbert space $\mathbb{H}_{\mathcal{S}}$, and denote the space of bounded, linear operators on $\mathbb{H}_{\mathcal{S}}$ by $\cL(\mathbb{H}_{\mathcal{S}})$. We now consider the notion of a frame, which generalizes the notion of basis \cite{Ferrie2008quasiprob, Ferrie2011quasiprob}. For our purposes, a frame $\mathcal{F}_\mathcal{S}$ for $\cL(\mathbb{H}_{\mathcal{S}}$) is any set $\mathcal{F}_\mathcal{S}:=\{M_{\boldsymbol{a}}\}_{\boldsymbol{a}}$ of Hermitian operators $M_{\boldsymbol{a}}$ that spans $\cL(\mathbb{H}_{\mathcal{S}})$. Such a (in general linearly-dependent) spanning set is sometimes referred to as over-complete basis of $\cL(\mathbb{H}_{\mathcal{S}})$. In turn, a frame $\mathcal{D}_\mathcal{S}:=\{\tilde{M}_{\boldsymbol{a}}\}_{\boldsymbol{a}}$ s.t.
 \begin{align}
    \mathcal{I} &=\sum_{\boldsymbol{a}}  \cket{\Tilde{M}_{\boldsymbol{a}}} \cbra{M_{\boldsymbol{a}}}
    \label{eq:frame_def},
  \end{align}
 where $\mathcal{I}$ is the identity map on $\cL(\mathbb{H}_{\mathcal{S}})$, is called dual to $\mathcal{F}_\mathcal{S}$ (and we then refer to $\mathcal{F}_\mathcal{S}$ as the primal to $\mathcal{D}_\mathcal{S}$). The round kets and bras appearing in Eq. \eqref{eq:frame_def} are a short-hand notation to denote operators in $\cL(\mathbb{H}_{\mathcal{S}})$ and their adjoints, respectively. Accordingly, we denote by $\cbraket{A}{B}$ the Hilbert-Schmidt inner product $\Tr[A^\dag B]$ in $\cL(\mathbb{H}_{\mathcal{S}})$. This is a popular notation in quantum information \cite{Caves1999cbra,Ferrie2008quasiprob, Ferrie2011quasiprob} that will used here interchangeably with the (more usual) operator notation upon convenience.

We take throughout $M_{\boldsymbol{a}}\geq0$ for all $\boldsymbol{a}$ and $\sum_{\boldsymbol{a}} {M_{\boldsymbol{a}}}=\openone_{\mathcal{S}}$, with $\openone_{\mathcal{S}}$ the identity operator on $\mathbb{H}_{\mathcal{S}}$, so that $\mathcal{F}_\mathcal{S}$ is a positive operator-valued measure (POVM) on $\mathbb{H}_{\mathcal{S}}$. POVMs  define generalized (i.e. beyond von Neumann) measurements \cite{Nielsen2011, Peres1995}. This, together with Eq. \eqref{eq:frame_def}, allows us to express any
 density operator $\varrho\in\cL(\mathbb{H}_{\mathcal{S}})$ as 
  \begin{align}
    \cket{\varrho} &=  
    \sum_{\boldsymbol{a}}  P_\varrho(\boldsymbol{a})\cket{\Tilde{M}_{\boldsymbol{a}}}, \label{eq:qci}
  \end{align}
  where $P_\varrho(\boldsymbol{a}):=\cbraket{M_{\boldsymbol{a}}}{\varrho}$ is the  probability of measurement outcome $\boldsymbol{a}$ on $\varrho$. 
 Eq. \eqref{eq:qci} is the basis of classical-shadow tomography, a powerful technique to get compact classical representations of states from measurements \cite{Aaronson2018shadtom, Huang2020shadow}.
 
Note that $M_{\boldsymbol{a}}\geq0$ for all $\boldsymbol{a}$ implies $\tilde{M}_{\boldsymbol{a}}\ngeq0$ in general \cite{Ferrie2008quasiprob, Ferrie2011quasiprob}. In addition, it will be useful to express the dual frame elements as affine combination of elements of $\mathcal{F}_\mathcal{S}$,
  \begin{align}
    \cket{\tilde{M}_{\boldsymbol{a}}} = \sum_{\boldsymbol{a}'} \mathfrak{T}_{\boldsymbol{a},\boldsymbol{a}'} \cket{M_{\boldsymbol{a}'}}, \forall\, \boldsymbol{a}
    \label{eq:cqi},
  \end{align}
 for some adequately chosen  $\mathfrak{T}$. With this parametrization, the primal- and dual-frame overlap matrices $T$ and $\tilde{T}$, respectively defined as $T_{\boldsymbol{a},\boldsymbol{a}'}:=\cbraket{M_{\boldsymbol{a}}}{M_{\boldsymbol{a}'}}$ and $\tilde{T}_{\boldsymbol{a},\boldsymbol{a}'}:=\cbraket{\tilde{M}_{\boldsymbol{a}}}{\tilde{M}_{\boldsymbol{a}'}}$, are related as $\tilde{T}= \mathfrak{T}\,T\, \mathfrak{T}$.
 
 An experimentally convenient choice of $\mathcal{F}_\mathcal{S}$ and $\mathcal{D}_\mathcal{S}$ is $M_{\boldsymbol{a}}=M_{a_1}\otimes \hdots \otimes M_{a_N}$ and $\tilde{M}_{\boldsymbol{a}}=\tilde{M}_{a_1}\otimes \hdots \otimes \tilde{M}_{a_N}$, for $\boldsymbol{a}:=(a_1,\hdots a_N)$. Here, $M_{a_j}$ is the $j$-th element of a single-qubit POVM frame and $\tilde{M}_{a_j}$ that of the corresponding dual frame. We refer to these as factorable frames. By virtue of Eqs. \eqref{eq:qci} and \eqref{eq:cqi}, these allow one to express any $\varrho$ as an affine combination of product states $\sigma_{\boldsymbol{a}}:=M_{\boldsymbol{a}}/t_{\boldsymbol{a}}$, where $t_{\boldsymbol{a}}:=\Tr[M_{\boldsymbol{a}}]$ \cite{Carrasquilla2019genrecon}. This fact has been used to reconstruct quantum states \cite{Carrasquilla2019genrecon}, processes \cite{Torlai2020qpt}, and overlaps \cite{Guerini2021stateoverlap} from single-qubit measurements. Additionally, this has been used to simulate quantum circuits \cite{Carrasquilla2021probsim} with generative machine learning models, where $\mathfrak{T}$ was taken as the canonical pseudo-inverse of $T$. However, other choices of $\mathfrak{T}$ are possible. It can be seen (see App. \ref{app:qpr}) that Eq. \eqref{eq:cqi} defines a dual to $\mathcal{F}_\mathcal{S}$ iff $\mathfrak{T}_{\boldsymbol{a},\boldsymbol{a}'}\in\mathbb{R}$, $\sum_{a} \mathfrak{T}_{\boldsymbol{a},\boldsymbol{a}'} = 1$, and 
  \begin{align}
    T &=T\, \mathfrak{T}\, T.\label{eq:t_tau_t}
  \end{align}
In general, the elements of $\mathfrak{T}$ can be positive or negative. As shown below, the negativity of $\mathfrak{T}$ governs the sample complexity of Monte Carlo estimations of expectation values of observables. Finally, note also that if $\mathfrak{T}$ fulfills Eq. \eqref{eq:t_tau_t}, necessarily so does $\tilde{T}= \mathfrak{T}\,T\, \mathfrak{T}$ ($\tilde{T}$ and $\mathfrak{T}$ collapsing to each other for the canonical choice of $\mathfrak{T}$ being a pseudo-inverse of $T$).

 \section{Interfaces for hybrid classical-quantum circuits \label{sec:quasiprobs}}
  
Our goal is to simulate quantum circuits using hybrid classical-quantum ones. More precisely, we are given an observable $O$, an $N$-qubit input state $\varrho_0:=\ketbra{0}{0}$, and a target circuit $C:=\{U_k\}_{k\in[f]}$, with $f\in\mathbb{N}$ single- or two-qubit unitary gates $U_k$. 
We denote by $s_k\subset\mathcal{S}$ the subset of 
qubits on which $U_k$ acts, and by $\boldsymbol{a}_{s_k}$ a corresponding sub-string of 
measurement outcomes on $s_k$. 
In addition, we use the short-hand notations $\overline{s}_k:=\mathcal{S}\setminus s_k$ for the 
qubits on which $U_k$ does not act and $\openone_{\overline{s}_k}$ for the identity on $\mathbb{H}_{\overline{s}_k}$. From the $f$ gates, $l<f$ are particularly experimentally demanding for NISQ implementations, and they are marked by the set of labels $L:=\{k_1,k_2,\hdots k_{l}\}$. The case we explicitly study below is that of two-qubit gates on qubits far apart in the connectivity graph in question. However, other relevant cases may be due to error mitigation convenience or other hardware-specific limitations, e.g. Either way, our goal is to estimate the expectation value $\Tr[\varrho_{f}\,O]$ of $O$ on the output state $\varrho_{f}:=U_{f}\hdots U_1\,\varrho_0\,U^{\dagger}_1\hdots U^{\dagger}_{f}$ by substituting every $U_k$ with $k\in L$  by a classical simulation of it. 

Our main tool to achieve this are interfaces between quantum objects and their (classical) frame representations. 
The first one is based on Eq. \eqref{eq:qci}. 
  \begin{definition}[\textit{Quantum-classical interfaces}]
  We refer as a QC interface on $s_k$ to the assignment of a classical snapshot $\Tilde{M}_{\boldsymbol{a}_{s_k}}$ to $s_k$ according to the measurement outcome $\boldsymbol{a}_{s_k}$ of a factorable POVM frame $\mathcal{F}_{s_k}$ on a state $\varrho\in\mathbb{H}_{\mathcal{S}}$, occurring with probability $P_\varrho(\boldsymbol{a}_{s_k})=(\openone_{\overline{s}_k}|(M_{\boldsymbol{a}_{s_k}}|\varrho)$.
  \end{definition}
  
 The second one is the reverse interface, which simulates $\Tilde{M}_{\boldsymbol{a}_{s_k}}$ as a linear combination of states $\sigma_{\boldsymbol{b}_{s_k}}:=M_{\boldsymbol{b}_{s_k}}/t_{\boldsymbol{b}_{s_k}}$. This is done by importance-sampling $\boldsymbol{b}_{s_k}$ from $\Tilde{T}^{(\mathcal{I}_{s_k})}$, given $\boldsymbol{a}_{s_k}$, with $\Tilde{T}^{(\mathcal{I}_{s_k})}$ the dual-frame overlap matrix on $s_k$. To see this, we apply on $|\Tilde{M}_{\boldsymbol{a}_{s_k}})$ the Hermitian conjugate of Eq. \eqref{eq:frame_def} and get $|\Tilde{M}_{\boldsymbol{a}_{s_k}})=\sum_{\boldsymbol{b}_{s_k}}\Tilde{T}^{(\mathcal{I}_{s_k})}_{\boldsymbol{a}_{s_k},\boldsymbol{b}_{s_k}}\, t_{\boldsymbol{b}_{s_k}}\,|\sigma_{\boldsymbol{b}_{s_k}})$. Then, using a standard trick, we rewrite  
  \begin{equation}
    \tilde{T}^{(\mathcal{I}_{s_k})}_{\boldsymbol{a}_{s_k},\boldsymbol{b}_{s_k}} =:\Big\|\tilde{T}^{(\mathcal{I}_{s_k})}_{\boldsymbol{a}_{s_k}}\Big\|_1\,P_{\mathcal{I}_{s_k}}(\boldsymbol{b}_{s_k}|\boldsymbol{a}_{s_k})\,\sgn\big(\tilde{T}^{(\mathcal{I}_{s_k})}_{\boldsymbol{a}_{s_k},\boldsymbol{b}_{s_k}}\big),\label{eq:prob_dist}
  \end{equation}
  where $\tilde{T}^{(\mathcal{I}_{s_k})}_{\boldsymbol{a}_{s_k}}$ is a short-hand notation for the vector given by the $\boldsymbol{a}_{s_k}$-th row of $\tilde{T}^{(\mathcal{I}_{s_k})}$, $\Big\|\tilde{T}^{(\mathcal{I}_{s_k})}_{\boldsymbol{a}_{s_k}}\Big\|_1:=\sum_{\boldsymbol{b}_{s_k}} \big|\tilde{T}^{(\mathcal{I}_{s_k})}_{\boldsymbol{a}_{s_k},\boldsymbol{b}_{s_k}}\big|$ its $l_1$-norm, and $P_{\mathcal{I}_{s_k}}(\boldsymbol{b}_{s_k}|\boldsymbol{a}_{s_k}):=\Big|\tilde{T}^{(\mathcal{I}_{s_k})}_{\boldsymbol{a}_{s_k},\boldsymbol{b}_{s_k}}\Big|/\Big\|\tilde{T}^{(\mathcal{I}_{s_k})}_{\boldsymbol{a}_{s_k}}\Big\|_1$. 
 By construction, $P_{\mathcal{I}_{s_k}}(\circ|\boldsymbol{a}_{s_k})$ is a valid probability distribution, from which $\boldsymbol{b}_{s_k}$ can be sampled. This can be used to quantum Monte Carlo simulate $\Tilde{M}_{\boldsymbol{a}_{s_k}}$ \cite{Pashayan2015quasiprobs}. 
  \begin{definition}[\textit{Classical-quantum interface}]
 We refer as CQ interface on $s_k$ to the repreparation of $s_k$ in the state $\sigma_{\boldsymbol{b}_{s_k}}$, with probability $P_{\mathcal{I}_{s_k}}(\boldsymbol{b}_{s_k}|\boldsymbol{a}_{s_k})$, given a classical snapshot $\Tilde{M}_{\boldsymbol{a}_{s_k}}$.  Each sampled duple $(\boldsymbol{a}_{s_k},\boldsymbol{b}_{s_k})$ is assigned the value $\Big\|\tilde{T}^{(\mathcal{I}_{s_k})}_{\boldsymbol{a}_{s_k}}\Big\|_1\, t_{\boldsymbol{b}_{s_k}}\,\sgn\big(\tilde{T}^{(\mathcal{I}_{s_k})}_{\boldsymbol{a}_{s_k},\boldsymbol{b}_{s_k}}\big)$.
  \end{definition}
%
 
 The third and final ingredient integrates QC and CQ interfaces with a classical simulation of $U_k$. Multiplying $\mathcal{U}_k$ from the right by Eq. \eqref{eq:frame_def} and from the left by the Hermitian conjugate of Eq. \eqref{eq:frame_def}, we get $\mathcal{U}_k=\sum_{\boldsymbol{a}_{s_k},\boldsymbol{b}_{s_k}} |M_{\boldsymbol{b}_{s_k}})\,\tilde{T}^{(\mathcal{U}_k)}_{\boldsymbol{b}_{s_k},\boldsymbol{a}_{s_k}}(M_{\boldsymbol{a}_{s_k}}|$, where $\tilde{T}^{(\mathcal{U}_k)}_{\boldsymbol{b}_{s_k},\boldsymbol{a}_{s_k}} := (\tilde{M}_{\boldsymbol{b}_{s_k}}|\,\mathcal{U}_k\,|\tilde{M}_{\boldsymbol{a}_{s_k}})$. With this, we get
  \begin{align}
    \mathcal{U}_k \cket{\varrho_{k-1}} &= \sum_{\boldsymbol{a},\boldsymbol{a}'} \tilde{T}^{(\mathcal{U}_k)}_{\boldsymbol{a}_{s_k},\boldsymbol{b}_{s_k}}\,t_{\boldsymbol{b}_{s_k}}\, |\sigma_{\boldsymbol{b}_{s_k}})\,(M_{\boldsymbol{a}_{s_k}}|\varrho_{k-1})\label{eq:uni_evo},
  \end{align}
where $\varrho_{k-1}=U_{k-1}\hdots U_1\,\varrho_0\,U^{\dagger}_1\hdots U^{\dagger}_{k-1}$. That is, the action of $\mathcal{U}_k$ is absorbed into the repreparation by sampling from $\tilde{T}^{(\mathcal{U}_k)}$ instead of $\tilde{T}^{(\mathcal{I}_{s_k})}$ (see Fig. \ref{fig:unitary}). This leads to:
  \begin{definition}[\textit{Quantum-classical-quantum interface}]
  \label{def:def3}
  We refer as a QCQ interface for $U_k$ on $s_k$ to the measurement of $\mathcal{F}_{s_k}$, with outcome $\boldsymbol{a}_{s_k}$, followed by the repreparation of $\sigma_{\boldsymbol{b}_{s_k}}$ with probability $P_{\mathcal{U}_k}(\boldsymbol{b}_{s_k}|\boldsymbol{a}_{s_k}):=\Big|\tilde{T}^{(\mathcal{U}_k)}_{\boldsymbol{a}_{s_k},\boldsymbol{b}_{s_k}}\Big|/\Big\|\tilde{T}^{(\mathcal{U}_k)}_{\boldsymbol{a}_{s_k}}\Big\|_1$. Each sampled duple $(\boldsymbol{a}_{s_k},\boldsymbol{b}_{s_k})$ is assigned the value $v_{\boldsymbol{a}_{s_k},\boldsymbol{b}_{s_k}}:=\Big\|\tilde{T}^{(\mathcal{U}_k)}_{\boldsymbol{a}_{s_k}}\Big\|_1\, t_{\boldsymbol{b}_{s_k}}\,\sgn\big(\tilde{T}^{(\mathcal{U}_k)}_{\boldsymbol{a}_{s_k},\boldsymbol{b}_{s_k}}\big)$; and the corresponding interface realized in such experimental run is thus mathematically represented by the operator $\mathcal{V}_k(\boldsymbol{a}_{s_k},\boldsymbol{b}_{s_k})\coloneqq v_{\boldsymbol{a}_{s_k},\boldsymbol{b}_{s_k}}\, |\sigma_{\boldsymbol{b}_{s_k}})(M_{\boldsymbol{a}_{s_k}}|$.
  \end{definition}

Our hybrid-circuit simulation then applies on $\varrho_{k-1}$ the gate $U_k$ if $k\notin L$, but a QCQ interface for $U_k$ instead if $k\in L$. Introducing the terminology
  \begin{align}
    \mathcal{W}_k(\boldsymbol{a}_{s_k},\boldsymbol{b}_{s_k})= 
\begin{cases} 
      \mathcal{U}_k, & \text{ if } k\notin L, \\
      \mathcal{V}_k(\boldsymbol{a}_{s_k},\boldsymbol{b}_{s_k}), & \text{ if } k\in L, 
       \end{cases}
\label{eq:hybrid_gate}
  \end{align}
and using the fact that $O$ is Hermitian, we can express the target expectation value $\Tr[\varrho_{f}\,O]$ as 
  \begin{equation}
   \cbraket{O}{\varrho_{f}} = \sum_{\boldsymbol{\alpha}_{s_L}}(O| \prod_{k=1}^{f}  \mathcal{W}_k(\boldsymbol{a}_{s_k},\boldsymbol{b}_{s_k}) |\varrho_{0})\label{eq:final_exp},
  \end{equation}
with the short-hand notation $\boldsymbol{\alpha}_{s_{L}}:=(\boldsymbol{a}_{s_{k_l}},\boldsymbol{b}_{s_{k_l}},\hdots\,\boldsymbol{a}_{s_{k_1}},\boldsymbol{b}_{s_{k_1}})$.
 Eq. \eqref{eq:final_exp} can be experimentally estimated through an average $\hat{O}_M$ over $M\in\mathbb{N}$ runs. $M$ is chosen to guarantee that the statistical error and significance level (failure probability) of the estimation are respectively given by target values $\varepsilon$ and $\delta$. We refer to $M$ as \emph{sample complexity} of the protocol and its explicit value is given in Theo. \ref{theorem1} below. 
 
 The procedure is sketched by the following pseudo-code.
 \RestyleAlgo{ruled}
  \begin{algorithm}[hbt!]
  \caption{Hybrid classical-quantum simulation with QCQ interfaces\label{alg:quasi}. }
  \KwIn{$\varrho_0$, $C$, $O$, $\varepsilon$, $\delta$}
  \KwOut{$\hat{O}_M$ s.t. $\big|\hat{O}_M-\text{Tr}\big[O\,\varrho_{f}\big]\big|\leq \varepsilon$ with probability at least $1-\delta$.}
  Initialize $\hat{O}_M = 0$, $v=1$, and $M$ as in Eq. \eqref{eq:M}.\\
  \For{$m \in (1, \ldots, M)$}{
  \For{$k \in (1, \ldots, f)$}{
    \eIf{$k \in L$}{
Apply a QCQ interface for $U_k$ on qubits $s_k$, obtaining the duple $(\boldsymbol{a}_{s_k},\boldsymbol{b}_{s_k})$;\\
$v\gets v\times v_{\boldsymbol{a}_{s_k},\boldsymbol{b}_{s_k}}$, with $v_{\boldsymbol{a}_{s_k},\boldsymbol{b}_{s_k}}$ as in Def. \ref{def:def3}.}
    {
    Apply the gate $U_k$ on qubits $s_k$.\\}
  }
 Measure $O$, obtaining the measurement outcome (eigenvalue of $O$) $o$;\\
 $\hat{O}_M \gets \hat{O}_M+o\times v$.}
 $\hat{O}_M \gets \frac{\hat{O}_M}{M}$.
  \end{algorithm}
    
To quantify the runtime of the algorithm, we define the \emph{interface negativity} of the gate $\mathcal{U}_{k}$ and  the \emph{total forward interface negativity} of the entire circuit $C$ respectively as 
\begin{equation}
\label{eq:def_neg}
n_{U_{k}}\coloneqq\max_{\boldsymbol{a}_{s_k},\boldsymbol{b}_{s_k}}\Big\|\tilde{T}^{(\mathcal{U}_{s_k})}_{\boldsymbol{a}_{s_k}}\Big\|_1\, t_{\boldsymbol{b}_{s_k}}\ \ \text{ and }\ \ n_{\to}\coloneqq\prod_{k\in L} n_{U_k}.
\end{equation} 
This allows us to state the following theorem. 
  \begin{theorem}
\label{theorem1}[\textit{Correctness and sample complexity}]
The finite-statistics average $O^*_M$ of Algorithm \ref{alg:quasi} is an unbiased estimator of $\Tr[\varrho_{f}\,O]$ (See App. \cref{app:finite_stats}). Moreover, if 
  \begin{equation}
  \label{eq:M}
    M\geq n_{\to}^{2}\times\frac{2\,\|O\|^2\,\log{(2/\delta)}}{\varepsilon^2}, 
  \end{equation}
with $\|O\|$ the operator norm of $O$, then, with probability at least $1-\delta$, the statistical error of $O^*_M$ is at most $\varepsilon$. 
  \end{theorem}
The proof follows straightforwardly from the Hoeffding bound. We note that the factor $\frac{2\,\|O\|^2\,\log{(2/\delta)}}{\varepsilon^2}$  in Eq. \eqref{eq:M} is the equivalent sample complexity bound one would obtain if $\Tr[\varrho_{f}\,O]$ was estimated from measurements on the actual state $\varrho_{f}$. Hence, $n_{\to}^{2}$ quantifies the runtime overhead introduced by the interfaces. In that regard, the interface negativities play the same role in our hybrid classical-quantum simulation as the negativities of Ref. \cite{Pashayan2015quasiprobs} in fully classical simulations with quasi-probability representations. An innovative and advantageous feature of Eq. \eqref{eq:def_neg} is the presence of the POVM-element trace $t_{\boldsymbol{b}_{s_k}}$ in $n_{U_{k}}$, which comes from the state repreparation. Indeed, since $t_{\boldsymbol{b}_{s_k}}<1$, the $n_{U_{k}}$'s (and therefore also $n_{\to}$) are significantly smaller than their counterparts for fully classical simulations \cite{Pashayan2015quasiprobs}. This is consistent with the intuition that hybrid classical-quantum Monte Carlo simulations should cause lower sample-complexity increases than fully classical ones.  

Either way, the most relevant property for our purposes is that $n_{\to}^{2}$ (and therefore also $M$) is independent not only of the numbers of gates $f$ or qubits $N$ but also, and most importantly, of the connectivity-graph distance between the qubits on which the interfaces act. In other words, for a fixed budget of measurement runs, simulating a gate $U_{k}$ with a QCQ interface increases the statistical error at most by a constant factor $n_{U_{k}}$, regardless how far apart in the circuit the qubits $s_k$ are. In contrast, experimentally synthesizing $U_{k}$ with noisy nearest-neighbor gates would give a systematic error due to infidelity accumulation that grows linearly with the distance between those qubits.   
Clearly, the drawback is that $n_{\to}^{2}$ grows exponentially with the number $l$ of interfaces used. However, for a many circuits, Alg. \ref{alg:quasi} constitutes a better alternative than the bare NISQ implementation.
We study relevant exemplary circuits with such trade-offs in the next sections. 

Finally, note that $n_{\to}^{2}$ is frame-dependent.  This is crucial to the efficiency of classical simulations  \cite{Hatano1992sign,Loh1990signprob,Troyer2005signprob}. For instance, in quantum Monte Carlo, it is known that the statistical overhead due to negative (quasi-)probabilities can be ameliorated \cite{Hangleiter2020easing} or even removed \cite{Marvian2019curing} by local base changes. Something similar applies here: the interface negativities depend not only on the primal frame but also on the choice of dual to it. 


  \section{Numerical experiments}\label{sec:numexp}
  Here, we provide numerical experiments to validate the procedure outlined in \Cref{alg:quasi}.
  Throughout the rest of this work, we take $\{M_{\boldsymbol{a}}\}_{\boldsymbol{a}}$ 
  to be the Pauli-6 Informationally Complete POVM (IC-POVM), which can be implemented in an experimental setting without the usage of ancilla qubits
  (see App. \ref{app:icpovm}). 
  For our simulations, we make use of full density matrix simulations and Locally Purified Density Operator tensor networks~\cite{Werner2016positivetensor} (see also App. \ref{app:lpdo}). For the latter, we choose the Kraus and Bond dimensions such that the simulation errors are under control and we end up with a high fidelity ($>99.9\%$) state approximation. To simulate realistic experimental settings, we apply noise to the two-qubit gates in our circuit. In particular we implement noisy CNOTs throughout our circuits by applying single-qubit depolarizing channels $\mathcal{E}: \varrho \mapsto \mathcal{E}(\varrho)$ to both the control and target qubit of the CNOT gate. We apply depolarizing noise in the CNOTs with $\lambda_{\text{unit}} = 0.005$. This values correspond to experimentally realistic values \cite{Arute2019syc}. At the end of the circuit we estimate observables $\Tr{\varrho O}$ exactly, i.e. without further sampling bitstrings but relying on the full state representation. 
To improve the sample complexity of our algorithm, we use a Monte Carlo algorithm to minimize the interface negativities. We first note that Eq. \eqref{eq:t_tau_t} defines a domain over which to optimize such negativity. Similar optimizations (but for bases instead of frames) have been used for alleviating  \cite{Marvian2019curing,Hangleiter2020easing} the sign problem in partition-function estimations.  In our setting, we use a convenient parametrization of generalized inverse matrices by Rao \cite{Rao1967geninv} to propose new dual frames for an adaptive random walk Metropolis-Hastings algorithm. This allows us to decrease the multiplicative sample complexity overhead per interface by almost a factor of four relative to the canonical dual frame (corresponding to $\mathfrak{T} = T^{-1}$, with $T^{-1}$ the pseudo-inverse of $T$), which reduces the number of samples required by a factor of four (see App. \ref{app:neg_min} for details).

\subsection{Simulation of long-range maximal Bell violations \label{sec:chsh}}
\begin{figure}[htb!]
    \includegraphics[width=\columnwidth]{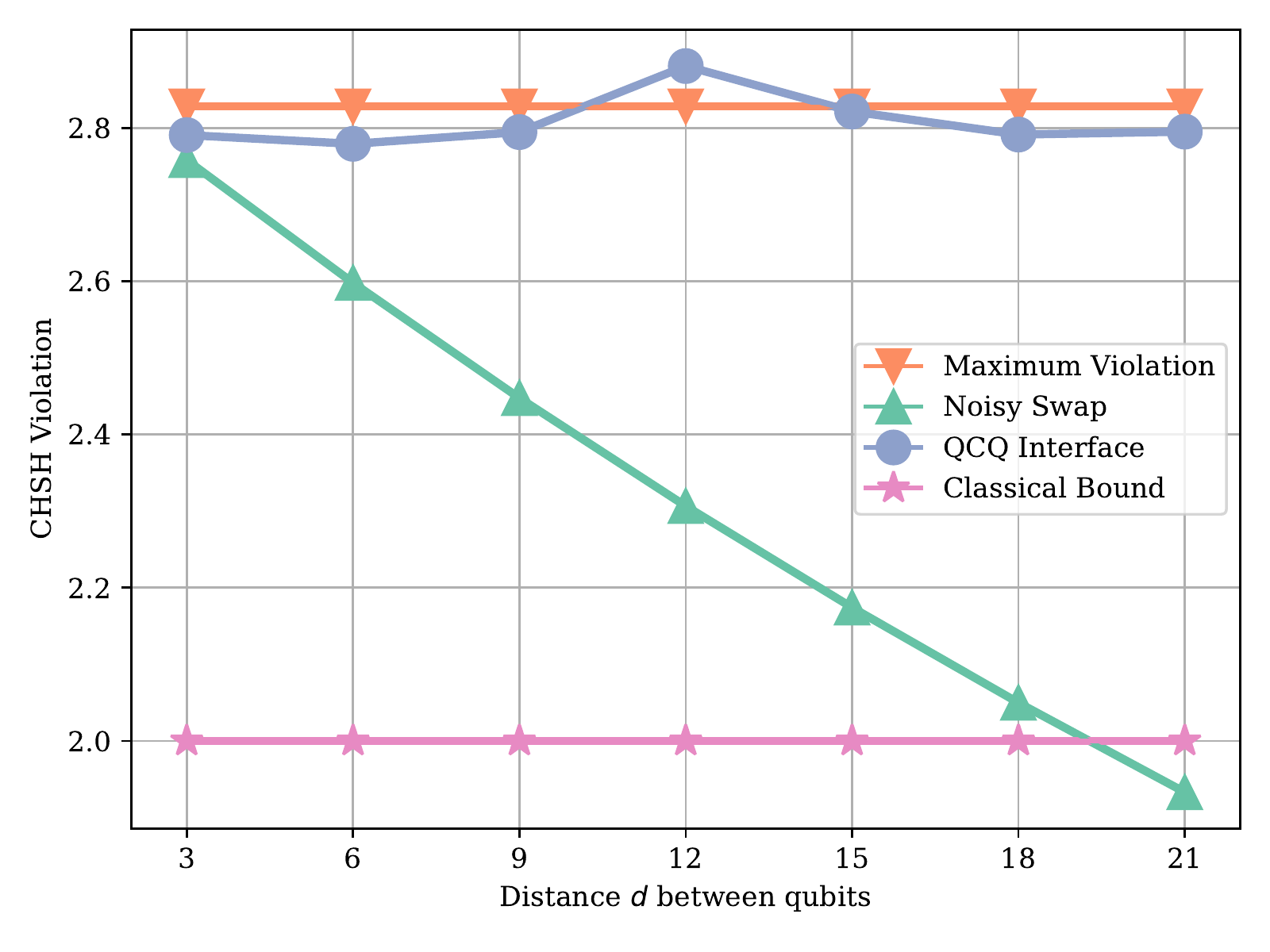}
    \caption{CHSH violation as a function the number of qubits. These results where obtained with an LPDO simulation where $D=12$ and $K=24$. In addition to the gate noise, we apply a depolarizing channel to simulate measurement noise with $\lambda_{\text{meas}}=0.01$ and repreparation noise with $\lambda_{\text{reprep}}=0.005$. The classical bound (pink) and maximal violation (green) are $2$ and $2\sqrt{2}$ respectively for all $d$. We see that the violation in the noisy circuit (green) decreases linearly with the number of qubits as a result of the $4(d-2)+1$ noisy swap gates required to prepare the state. Our algorithm provides the maximum CHSH violation up to statistical fluctuations independent of the distance between the qubits. This comes at a cost of sampling $M=60000$ measurement-and-reprepare steps to estimate the violation.
    }
    \label{fig:chsh}
  \end{figure}

  As a proof of principle experiment, we show that a maximally entangled state simulated with our method attains the maximal violation of the Clauser-Horne-Shimony-Holt (CHSH) inequalities (see App. \ref{app:chsh}) as expected. Specifically, we create the Bell state $\ket{\Phi^+} = \frac{1}{2}(\ket{00}+\ket{11})$ which has the maximum CHSH violation $S(A,B) = 2\sqrt{2}$. We consider the case where the state is prepared on two qubits separated by a distance $d$. Applying the CNOT between these distant qubits requires implementing a swap chain to bring the two states close together. In Fig. \ref{fig:chsh} we compare the CHSH violation of the Bell state simulated with our algorithm and one prepared with a circuit containing a noisy swap chain. We see that the CHSH violation is only affected by the statistical fluctuations of our method and therefore approximates the maximum value independent of the distance between the qubits. 

  
  
%

  \subsection{The Transverse Field Ising-model circuit}
  \label{sec:TFIsing}
\begin{figure}[htb!]
    \includegraphics[width=0.9\columnwidth]{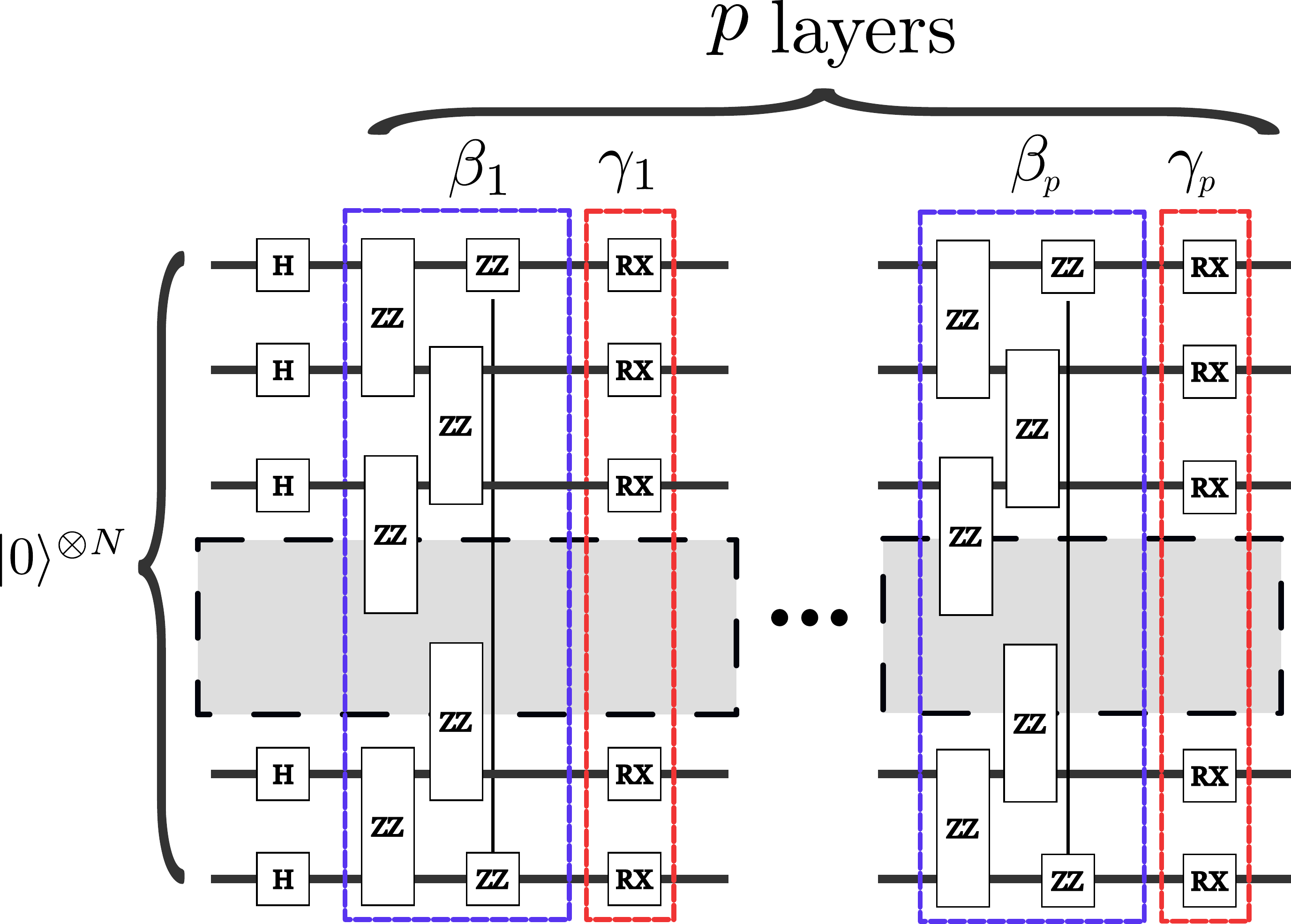}
    \caption{The Hamiltonian Variational Ansatz  circuit for the ground state of the TFIM model. The parameters $\{\beta_i,\gamma_i\}$ for $i=1,\ldots,p$ can be found with a Variational Quantum Eigensolver optimization \cite{Peruzzo2014vqe}. Each layer in the circuit contains a long range two qubit ZZ rotation. We assume that the distance between the first and last qubit is $N-2$. Implementing the nearest neighbor ZZ gates come at a cost of $2(N-1)$ CNOTs. The long range ZZ rotations require $4(N-2)+1$ CNOTs since we must use a swap chain to bring the first and last qubit together. The total number of CNOT gates per layer is therefore dominated by the implementation of long range ZZ.}
    \label{fig:tfim_circuit}
  \end{figure}

 As a practical example of implementing our method in an experimentally realistic setting, we investigate the ground state of a prototypical model for quantum magnetism: the transverse field Ising-model (TFIM) on a one dimensional ring. The Hamiltonian of the TFIM for the 1D chain is given by
  \begin{equation}\label{eq:tfim}
      H_{\text{TFIM}} = -\sum_{i=1}^N\left[ Z_i Z_{i+1} + g X_i \right]
  \end{equation}
  where we assume periodic boundary conditions and set $g=1$. The ground state of $H$ can be approximated reliably with a depth $p=N/2$ circuit ansatz called the Hamiltonian Variational Ansatz  \cite{Ho2019hva, Wierichs2020avoiding, Wiersema2020exploring}.
  This circuit for the ground state is given in Fig. \ref{fig:tfim_circuit}.
   To evaluate the accuracy of the state reconstruction, we compare the finite statistics estimator of the energy $\expval{\hat{H}_M}$ from our algorithm with the ground-state energy $E_{\text{gs}}=\bra{\psi_{\text{gs}}}H\ket{\psi_{\text{gs}}}$ from exact diagonalization. 
   
\begin{figure*}[htb!]
    \subfloat[\label{fig:tfim_4_single} 4-qubit TFIM]{%
    \centering
    \includegraphics[width=\columnwidth]{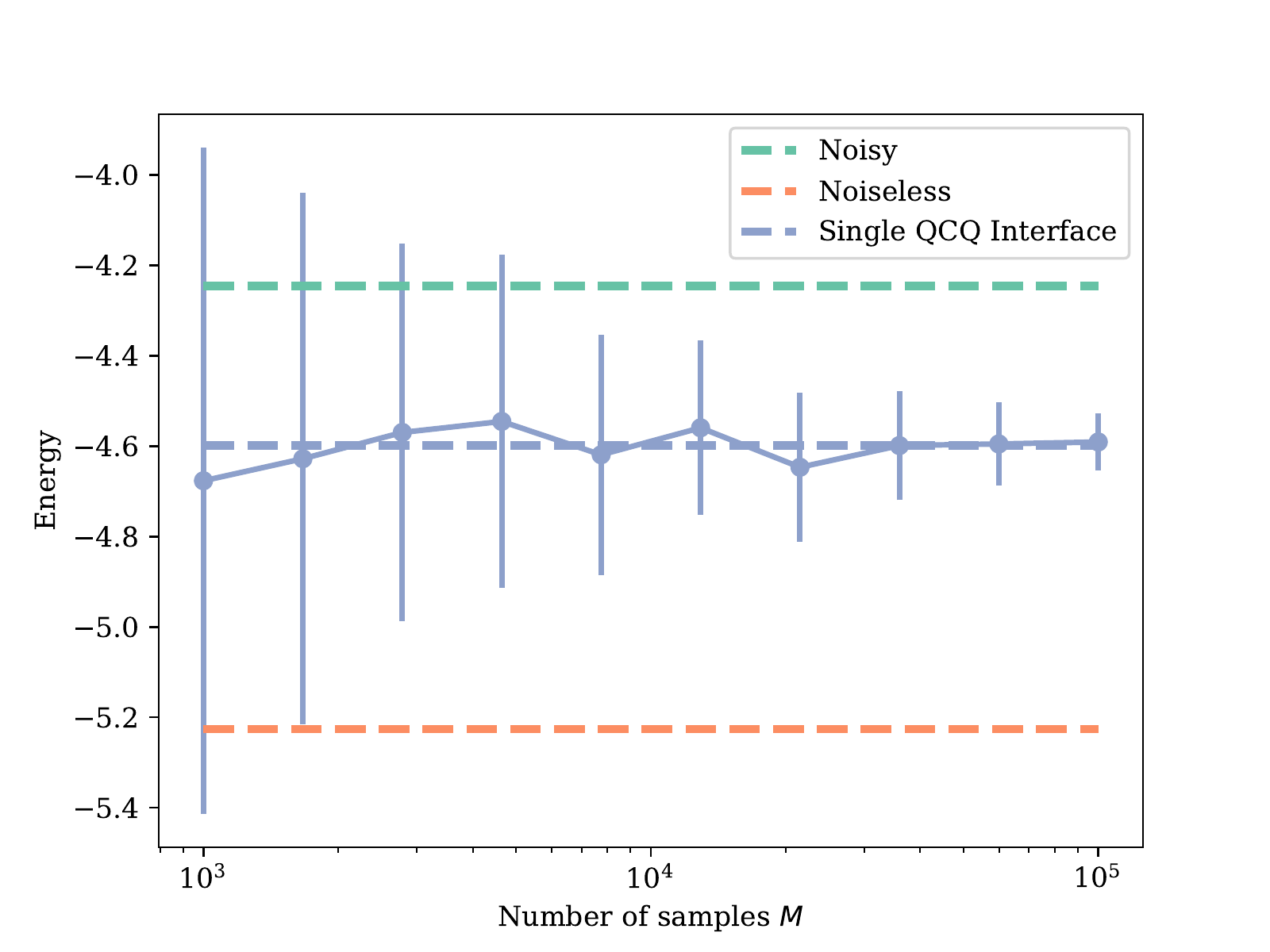}
  }
  \subfloat[\label{fig:tfim_8_single} 8-qubit TFIM]{%
    \centering
    \includegraphics[width=\columnwidth]{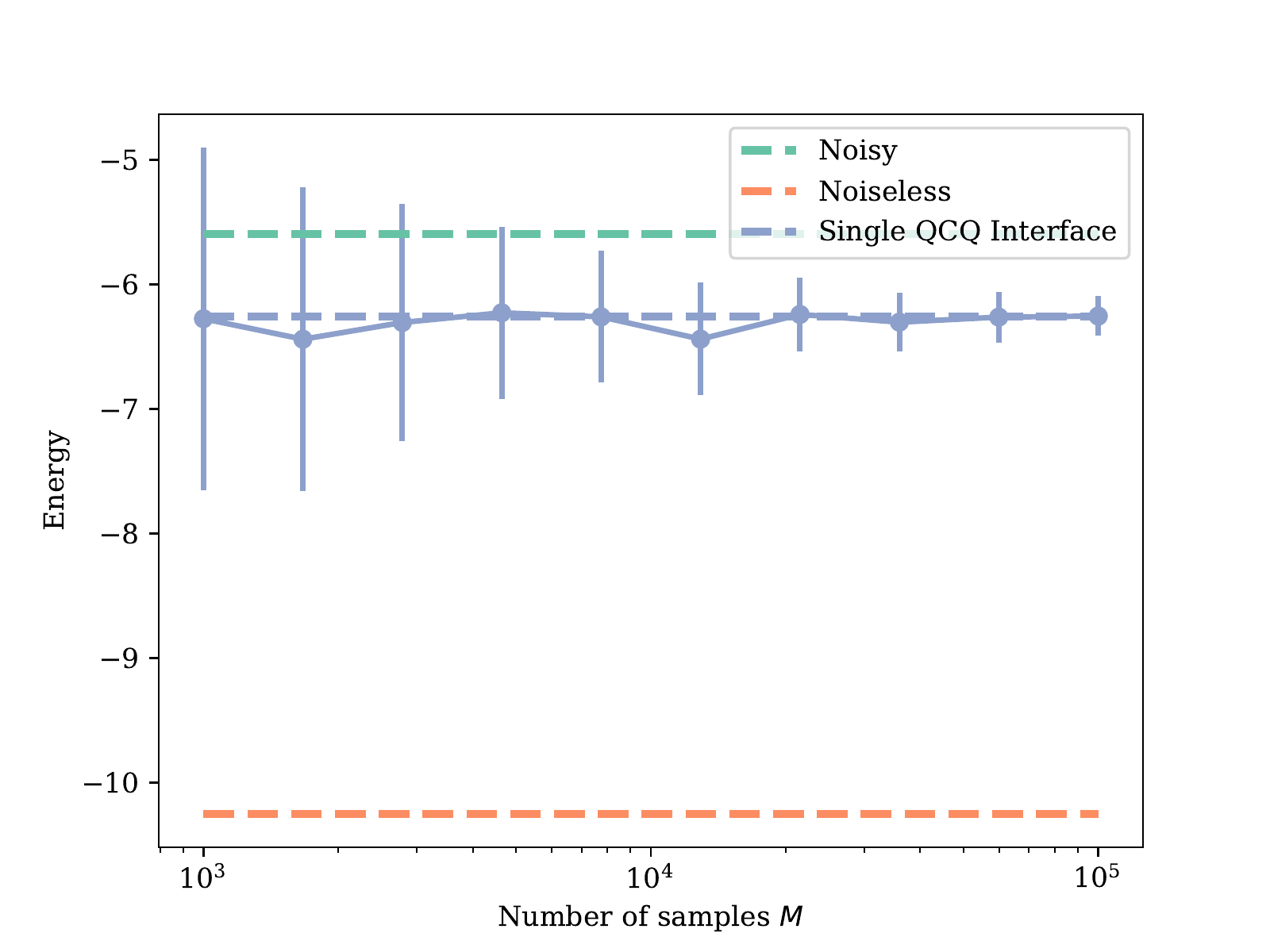}
  }
  \caption{Comparison of QCQ interface simulation with both noisy and noiseless TFIM circuits for (a) $N=4$ and (b) $N=8$ qubits obtained with a full density matrix simulation. Each dot represents the average energy $\mathbb{E}\left[\expval{\hat{H}_M}\right]$ estimated over 50 separate instances. The error bars indicate the standard deviation. As the number of samples $M$ increases, the statistical fluctuations of our method become small in accordance with the central limit theorem. We can determine the scaling of the size of the error bars by fitting $\sigma = \bar{\sigma} / \sqrt{N_{\text{samples}}}$. While for 4 qubits $\bar{\sigma} \approx 27.8 $, for 8 qubits we have $\bar{\sigma} \approx 76.5$. This scaling only depends on the mean negativity, which differs between the two circuits because we apply a different ZZ rotation on each circuit. The energy of the noiseless circuit (orange dashed line) corresponds to the ground-state energy $E_{\text{gs}}$. The noisy circuit (green dashed line) shows the energy obtained when we apply depolarizing channels with $\lambda_{\text{unit}} = 0.005$ to the CNOT gates in the circuit. We see that for both the 4 and 8 qubit our algorithm provides a significant improvement on the final estimated energy of the circuit for a reasonable number of measure-and-reprepare steps. In (b) we observe that the large number of number of noisy CNOTs dominates the simulation, hence the improvement is not as significant as for 4 qubit.
}
  \label{fig:tfim_n_single}
  \end{figure*}

\begin{figure*}[htb!]
    \subfloat[\label{fig:tfim_4_twice} 4 qubit TFIM]{%
    \centering
    \includegraphics[width=\columnwidth]{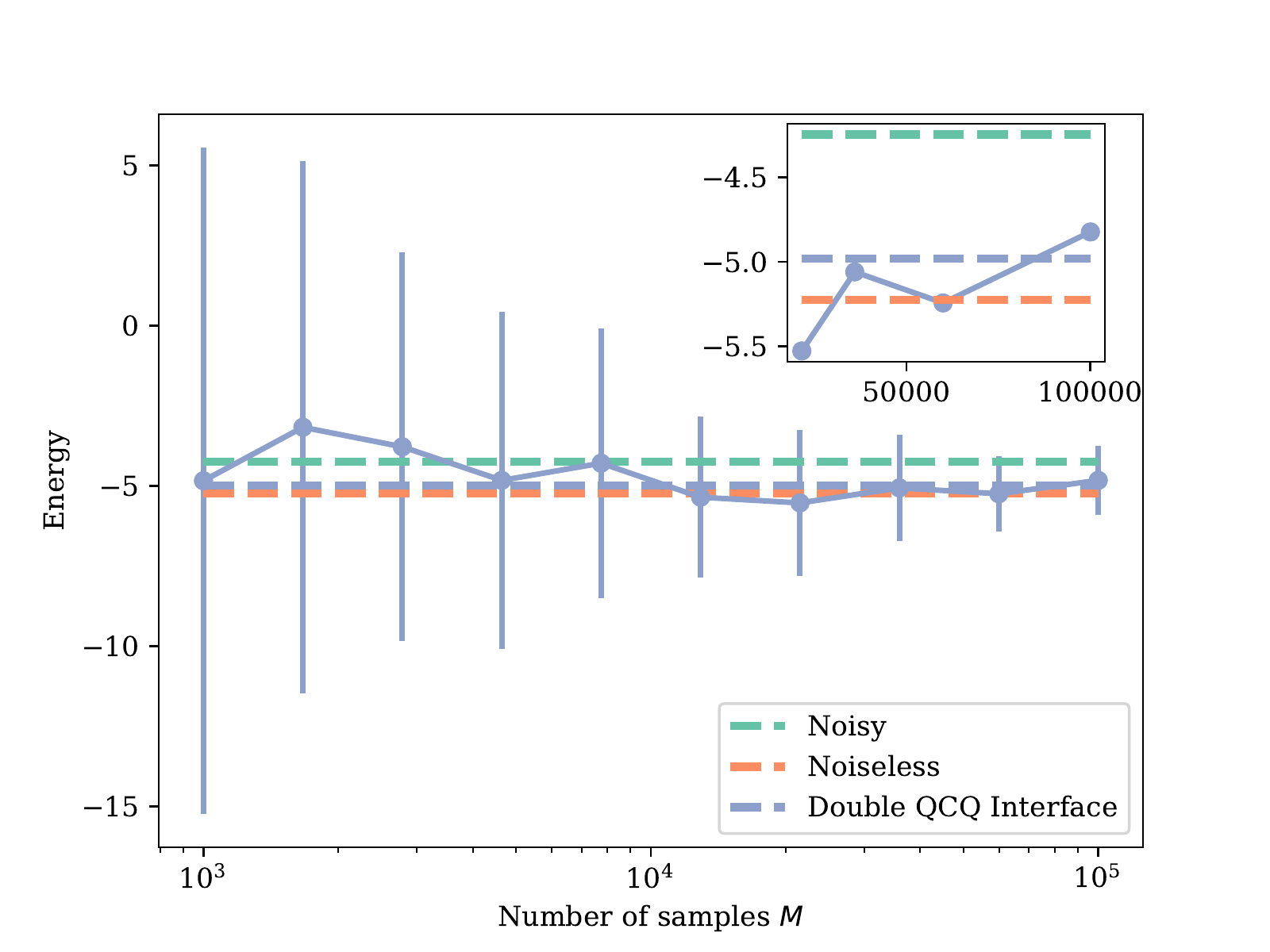}  }
    \subfloat[\label{fig:tfim_8_twice} 8 qubit TFIM]{%
    \centering
    \includegraphics[width=\columnwidth]{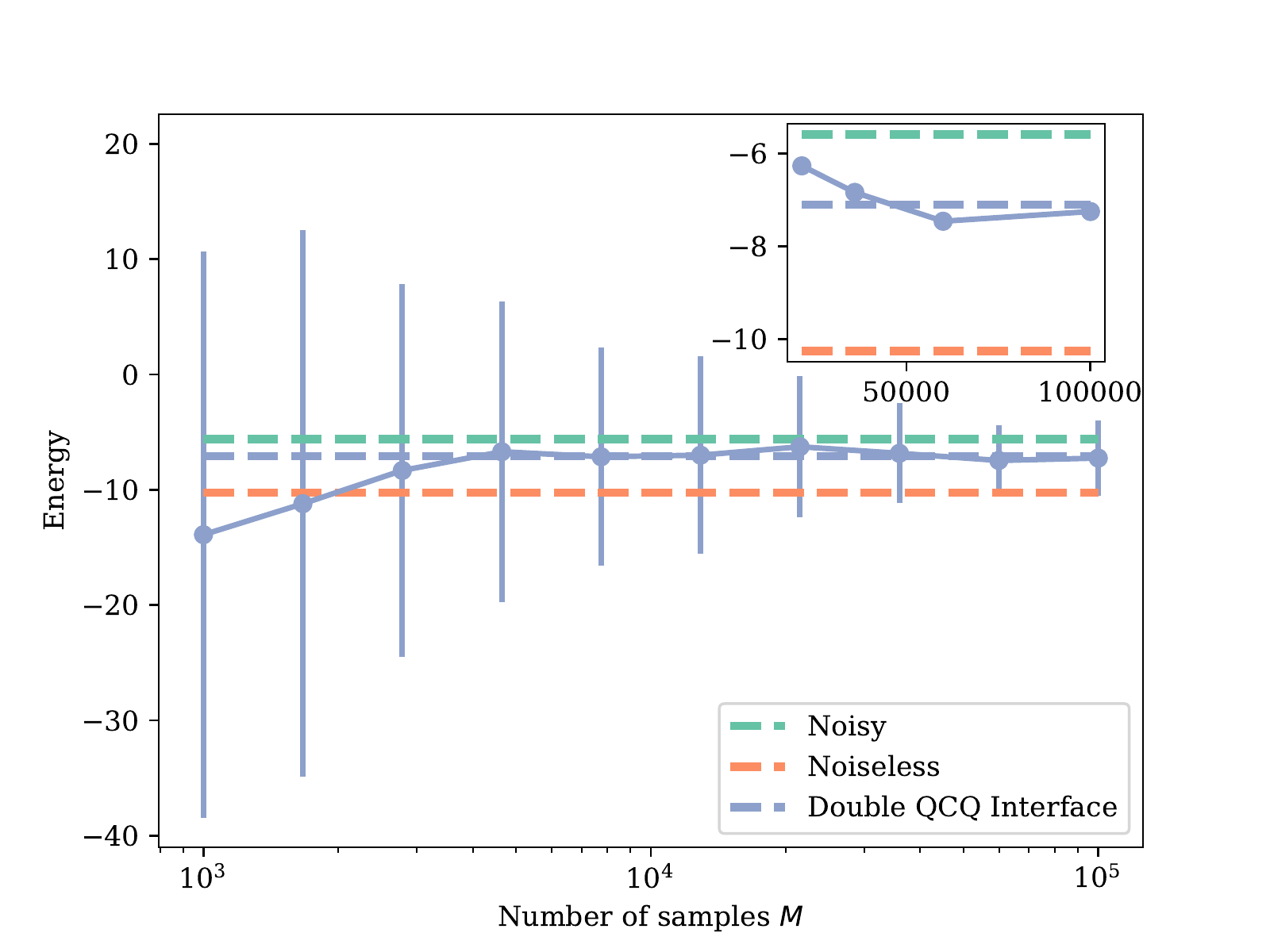}
  }
  \caption{Comparison of double QCQ interface simulation with both noisy and noiseless TFIM circuits for (a) $N=4$ and (b) $N=8$ qubits. These results were obtained with a full density matrix simulation. In (a), we see that we can almost approximate the true ground-state energy of the 4 qubit state, because the only noisy operations are the 12 CNOTs required for implementing the 6 nearest neighbor ZZ gates in layers 1 and 2. In (b) we see a more significant improvement over the energies from \Cref{fig:tfim_8_single}, but still the noise dominates. Since we apply the QCQ method twice, the standard deviation $\sigma = \bar{\sigma} / \sqrt{N_{\text{samples}}}$ of the error bars increases quadratically, as per Eq. \eqref{eq:def_neg}. We find $\bar{\sigma}\approx333.1$ and $\bar{\sigma}\approx 856.8$ for 4 and 8 qubits respectively.
  }
  \label{fig:tfim_n_twice}
  \end{figure*}

\begin{figure}[htb!]
    \includegraphics[width=\columnwidth]{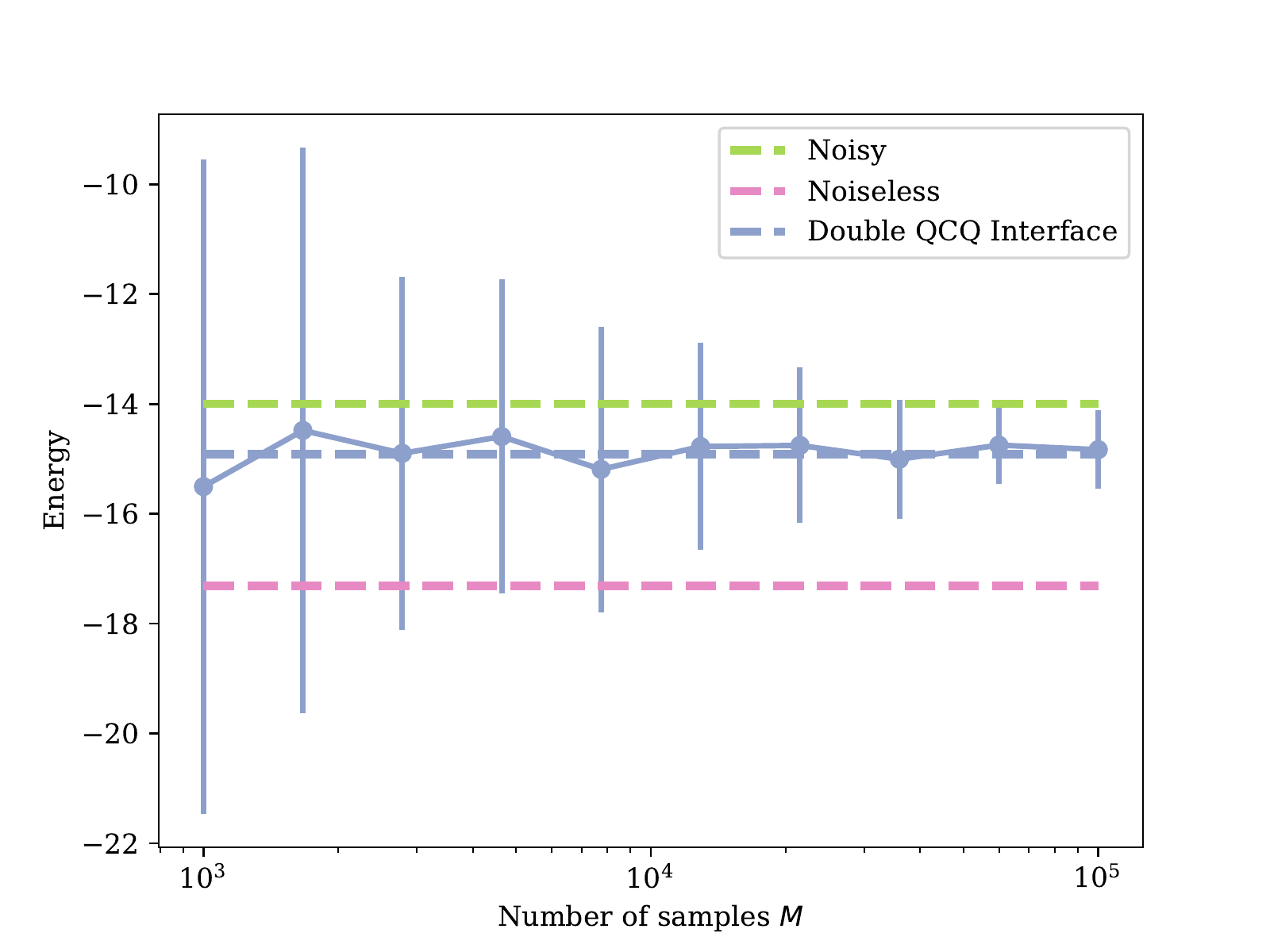}
    \caption{Comparison of a QCQ interface simulation with both noisy and noiseless circuits for a 20 qubit TFIM circuit. These results where obtained with an LPDO simulation where $D=50$ and $K=50$. Only two of the 8 layers of the circuit are simulated here, to keep simulation errors under control. The sample variance $\hat{\sigma} \approx 195.0$.
    }
    \label{fig:tfim_20_single}
  \end{figure}

   We consider three setups: First, we consider the $N=4$ and $N=8$ qubit TFIM chains where the last long range ZZ gate (in the 2nd and 4th layer respectively) is classically simulated with our algorithm (See Fig. \ref{fig:tfim_n_single}). Next, we apply our method twice for the same circuits, with simulation of both the last and first-to-last long range ZZ gate (See Fig. \ref{fig:tfim_n_twice}). Finally, we consider the ground state of a $N=20$ TFIM chain, where we only apply the first two layers of the circuit and simulate the second long range ZZ gate (See Fig. \ref{fig:tfim_20_single}). For all experiments, we confirm that we can greatly improve the final energy estimates by making use of QCQ interfaces at the cost of $M$ measurement-and-reprepare steps. 
   

  \section{Final discussion}\label{sec:final}
We have introduced a rigorous framework of hybrid quantum-classical interfaces for quantum-circuit simulations. 
We applied a specific variant of these gadgets -- which we dub quantum-classical-quantum (QCQ) interfaces -- to simulate long-range gates in low-connectivity devices without using swap-gate ladders. QCQ interfaces replaces an experimentally problematic gate (e.g. a very long-range one) by single-qubit random measurements and state-preparations sampled according to a classical quasi-probability simulation of the ideal target gate. This procedure eliminates long swap-gate ladders which would otherwise be required to physically synthesize the target gate. This results in a drastic increase in gate fidelity. The final output of the scheme is an estimate of the expectation value of a given observable on the output of the target high-connectivity circuit.

The quasi-probability distribution used is given by a frame representation of the gate simulated at each interface. As any sampling scheme based on non-positive quasi-probabilities, our method suffers from the sign problem. Because of this, the overall sample complexity grows exponentially with the number of interfaces applied. However, the statistical overhead per interface is independent of the on-chip distance between the qubits on which the interface acts. To ameliorate the sign problem, we developed a Metropolis-Hastings simulated-annealing algorithm based on random walks in the space of dual frames. This allowed us to decrease the statistical overhead per interface by almost a factor of two over that of the canonical dual frame. This is potentially interesting on its own beyond the current scope and further optimization is possible. All together, we show that any circuit with a limited number of gates to cut out can be simulated at the expenses of a moderate overall overhead in sample complexity. 
As examples, we explicitly considered a Bell-state preparation circuit for two qubits increasingly far apart and variational ground-state solvers for the transverse-field Ising model on ring lattices. The former involves a single long-range gate, whereas the latter one such gate per variational layer. 

Importantly, our method requires platforms supporting mid-circuit measurements and state preparations, which are readily provided by some quantum hardware companies such as, e.g., IBM and Honeywell \cite{Honeywell, IBM}. This may pave the way to implement our method in a practical setting in the near future.

Finally, we emphasize that our framework is not restricted to connectivity boosts only. It could also be applied to any gate that is too noisy for a given platform or combined with error-correcting codes to remove a gate that is particularly difficult to implement fault-tolerantly by the code. Another interesting application that will be studied elsewhere is circuit-depth boosts, where a deep circuit is simulated by shallower experimental circuits together with classical simulations of entire slices of the target circuit.  In conclusion, our framework provides a versatile toolbox for both error-mitigation and circuit boots well suited for noisy, intermediate-scale quantum hardware.
    
  \section{Acknowledgements}
  The authors thank Ingo Roth for
helpful insights. LG and LA acknowledge support by the Serrapilheira Institute (grant number Serra1709- 17173) and the Brazilian agencies CNPq (PQ grant No. 311416/2015-2 and INCT-IQ), FAPERJ (PDR10
E- 26/202.802/2016 and JCN E-26/202.701/2018), and FAPESP (grants
2016/01343-7 and 2018/04208-9). JC acknowledges support from the Natural Sciences and Engineering Research
Council of Canada (NSERC), the Shared Hierarchical
Academic Research Computing Network (SHARCNET),
Compute Canada, Google Quantum Research Award,
and the CIFAR AI chair program. Resources used in
preparing this research were provided, in part, by the
Province of Ontario, the Government of Canada through
CIFAR, and companies sponsoring the Vector Institute
\url{www.vectorinstitute.ai/\#partners}.

  \bibliographystyle{unsrt}
  \bibliography{biblio.bib}

\begin{thebibliography}{10}

\bibitem{Q_alg_review_Montanaro}
A.~Montanaro.
\newblock {Quantum algorithms: an overview}.
\newblock {\em NPJ Quantum Information}, 2:15023, 2016.

\bibitem{Q_Machine_Learning_rev}
J.~Biamonte, P.~Wittek, N.~Pancotti, P.~Rebentrost, N.~Wiebe, and S.~Lloyd.
\newblock {Quantum machine learning}.
\newblock {\em Nature}, 549:195--202, 2017.

\bibitem{Q_Finance_review}
A.~Bouland, W.~van Dam, H.~Joorati, I.~Kerenidis, and A.~Prakash.
\newblock {Prospects and challenges of quantum finance}.
\newblock {\em ArXiv preprint: 2011.06492}, 2021.

\bibitem{Preskill2018nisq}
John Preskill.
\newblock {Quantum {C}omputing in the {NISQ} era and beyond}.
\newblock {\em {Quantum}}, 2:79, Aug 2018.

\bibitem{Par_Q_circuits19}
M.~Benedetti, E.~Lloyd, S.~Sack, and M.~Fiorentini.
\newblock {Parameterized quantum circuits as machine learning models}.
\newblock {\em Quantum Science and Technology}, 4:043001, 2019.

\bibitem{NISQ_review}
K.~Bharti, A.~Cervera-Lierta, T.~H. Kyaw, T.~Haug, S.~Alperin-Lea, A.~Anand,
  M.~Degroote, H.~Heimonen, J.~S. Kottmann, T.~Menke, W.-K. Mok, S.~Sim, L.-C.
  Kwek, and A.~Aspuru-Guzik.
\newblock {Noisy intermediate-scale quantum (NISQ) algorithms}.
\newblock {\em ArXiv preprint: 2101.08448}, 2021.

\bibitem{Variational_q_algs_review}
M.~Cerezo, A.~Arrasmith, R.~Babbush, S.~C. Benjamin, S.~Endo, K.~Fujii, J.~R.
  McClean, K.~Mitarai, X.~Yuan, L.~Cincio, and P.~J. Coles.
\newblock {Variational quantum algorithms}.
\newblock {\em Nature Reviews Physics}, 3(625-644), 2021.

\bibitem{QAOA}
E.~Farhi, J.~Goldstone, and S.~Gutmann.
\newblock {A Quantum Approximate Optimization Algorithm}.
\newblock {\em ArXiv preprint: 1411.4028}, 2014.

\bibitem{Peruzzo2014vqe}
Alberto Peruzzo, Jarrod McClean, Peter Shadbolt, Man-Hong Yung, Xiao-Qi Zhou,
  Peter~J. Love, Al{\'a}n Aspuru-Guzik, and Jeremy~L. O'Brien.
\newblock {A variational eigenvalue solver on a photonic quantum processor}.
\newblock {\em Nature Communications}, 5(1):4213, Jul 2014.

\bibitem{Temme2017errormit}
Kristan Temme, Sergey Bravyi, and Jay~M. Gambetta.
\newblock {Error Mitigation for Short-Depth Quantum Circuits}.
\newblock {\em Phys. Rev. Lett.}, 119:180509, Nov 2017.

\bibitem{Endo2018errormit}
Suguru Endo, Simon~C. Benjamin, and Ying Li.
\newblock {Practical Quantum Error Mitigation for Near-Future Applications}.
\newblock {\em Phys. Rev. X}, 8:031027, Jul 2018.

\bibitem{Yi2017activeerror}
Ying Li and Simon~C. Benjamin.
\newblock {Efficient Variational Quantum Simulator Incorporating Active Error
  Minimization}.
\newblock {\em Phys. Rev. X}, 7:021050, Jun 2017.

\bibitem{Kandala2019errormit}
Abhinav Kandala, Kristan Temme, Antonio~D. C{\'o}rcoles, Antonio Mezzacapo,
  Jerry~M. Chow, and Jay~M. Gambetta.
\newblock {Error mitigation extends the computational reach of a noisy quantum
  processor}.
\newblock {\em Nature}, 567(7749):491--495, Mar 2019.

\bibitem{Bravyi2016trading}
Sergey Bravyi, Graeme Smith, and John~A. Smolin.
\newblock {Trading Classical and Quantum Computational Resources}.
\newblock {\em Phys. Rev. X}, 6:021043, Jun 2016.

\bibitem{Dunjko18}
V.~Dunjko, Y.~Ge, and J.~I. Cirac.
\newblock {Computational speedups using small quantum devices}.
\newblock {\em Phys. Rev. Lett.}, 121:250501, 2018.

\bibitem{Peng2020smallqc}
Tianyi Peng, Aram~W. Harrow, Maris Ozols, and Xiaodi Wu.
\newblock {Simulating Large Quantum Circuits on a Small Quantum Computer}.
\newblock {\em Phys. Rev. Lett.}, 125:150504, Oct 2020.

\bibitem{Tang2021circuitcut}
Wei Tang, Teague Tomesh, Martin Suchara, Jeffrey Larson, and Margaret
  Martonosi.
\newblock {CutQC: Using Small Quantum Computers for Large Quantum Circuit
  Evaluations}.
\newblock In {\em Proceedings of the 26th ACM International Conference on
  Architectural Support for Programming Languages and Operating Systems},
  ASPLOS 2021, pages 473--486, New York, NY, USA, 2021. Association for
  Computing Machinery.

\bibitem{Perlin2021circuitcut}
Michael~A. Perlin, Zain~H. Saleem, Martin Suchara, and James~C. Osborn.
\newblock {Quantum circuit cutting with maximum-likelihood tomography}.
\newblock {\em NPJ Quantum Information}, 7(1):64, Apr 2021.

\bibitem{Carrasquilla2019genrecon}
Juan Carrasquilla, Giacomo Torlai, Roger~G. Melko, and Leandro Aolita.
\newblock {Reconstructing quantum states with generative models}.
\newblock {\em Nature Machine Intelligence}, 1(3):155--161, Mar 2019.

\bibitem{Huang2020shadow}
Hsin-Yuan Huang, Richard Kueng, and John Preskill.
\newblock {Predicting many properties of a quantum system from very few
  measurements}.
\newblock {\em Nature Physics}, 16(10):1050--1057, Oct 2020.

\bibitem{Ferrie2008quasiprob}
Christopher Ferrie and Joseph Emerson.
\newblock {Frame representations of quantum mechanics and the necessity of
  negativity in quasi-probability representations}.
\newblock {\em Journal of Physics A: Mathematical and Theoretical},
  41(35):352001, Jul 2008.

\bibitem{Ferrie2011quasiprob}
Christopher Ferrie.
\newblock {Quasi-probability representations of quantum theory with
  applications to quantum information science}.
\newblock {\em Reports on Progress in Physics}, 74(11):24, 2011.

\bibitem{Mari_Eisert12}
A.~Mari and J.~Eisert.
\newblock {Positive Wigner functions render classical simulation of quantum
  computation efficient}.
\newblock {\em Phys. Rev. Lett.}, 109:230503, 2012.

\bibitem{Veitch_12}
V.~Veitch, N.~Wiebe, C.~Ferrie, and J.~Emerson.
\newblock {Efficient simulation scheme for a class of quantum optics
  experiments with non-negative {W}igner representation}.
\newblock {\em New Journal of Physics}, 15:013037, 2013.

\bibitem{Pashayan2015quasiprobs}
Hakop Pashayan, Joel~J. Wallman, and Stephen~D. Bartlett.
\newblock {Estimating Outcome Probabilities of Quantum Circuits Using
  Quasiprobabilities}.
\newblock {\em Physical Review Letters}, 115(7), Aug 2015.

\bibitem{Hatano1992sign}
Naomichi Hatano and Masuo Suzuki.
\newblock {Representation basis in quantum Monte Carlo calculations and the
  negative-sign problem}.
\newblock {\em Physics Letters A}, 163(4):246--249, 1992.

\bibitem{Loh1990signprob}
E.Y Loh, J.E Gubernatis, R.T Scalettar, S.R White, D.J Scalapino, and R.L
  Sugar.
\newblock {Sign problem in the numerical simulation of many-electron systems}.
\newblock {\em Physical review. B, Condensed matter}, 41(13):9301--9307, 1990.

\bibitem{Troyer2005signprob}
Matthias Troyer and Uwe-Jens Wiese.
\newblock {Computational Complexity and Fundamental Limitations to Fermionic
  Quantum Monte Carlo Simulations}.
\newblock {\em Phys. Rev. Lett.}, 94:170201, May 2005.

\bibitem{Rao1967geninv}
C.~Radhakrishna Rao.
\newblock {Calculus of Generalized Inverses of Matrices Part I: General
  Theory}.
\newblock {\em Sankhy{\=a}: The Indian Journal of Statistics, Series A
  (1961-2002)}, 29(3):317--342, 1967.

\bibitem{Marvian2019curing}
Milad Marvian, Daniel~A. Lidar, and Itay Hen.
\newblock {On the computational complexity of curing non-stoquastic
  Hamiltonians}.
\newblock {\em Nature Communications}, 10(1):1571, Apr 2019.

\bibitem{Hangleiter2020easing}
D.~Hangleiter, I.~Roth, D.~Nagaj, and J.~Eisert.
\newblock {Easing the Monte Carlo sign problem}.
\newblock {\em Science advances}, 6(33):eabb8341, 2020.

\bibitem{Caves1999cbra}
Carlton~M. Caves.
\newblock {Quantum Error Correction and Reversible Operations}.
\newblock {\em Journal of Superconductivity}, 12(6):707--718, Dec 1999.

\bibitem{Nielsen2011}
Michael~A. Nielsen and Isaac~L. Chuang.
\newblock {\em {Quantum Computation and Quantum Information}}.
\newblock Cambridge University Press, USA, 10th edition, 2011.

\bibitem{Peres1995}
Asher Peres.
\newblock {\em {Quantum theory : concepts and methods}}.
\newblock Fundamental theories of physics. Kluwer Academic, Dordrecht, 1993.

\bibitem{Aaronson2018shadtom}
Scott Aaronson.
\newblock Shadow tomography of quantum states.
\newblock In {\em {Proceedings of the 50th Annual ACM SIGACT Symposium on
  Theory of Computing}}, STOC 2018, pages 325--338, New York, NY, USA, 2018.
  Association for Computing Machinery.

\bibitem{Torlai2020qpt}
Giacomo Torlai, Christopher~J. Wood, Atithi Acharya, Giuseppe Carleo, Juan
  Carrasquilla, and Leandro Aolita.
\newblock {Quantum process tomography with unsupervised learning and tensor
  networks}.
\newblock {\em ArXiv preprint: 2006.02424}, 2020.

\bibitem{Guerini2021stateoverlap}
Juan Felipe~Carrasquilla Leonardo~Guerini, Roeland~Wiersema and Leandro Aolita.
\newblock {Quasiprobabilistic state-overlap estimator for NISQ devices}.
\newblock {\em ArXiv preprint: 2112.11618}, 2022.

\bibitem{Carrasquilla2021probsim}
Juan Carrasquilla, Di~Luo, Felipe P\'erez, Ashley Milsted, Bryan~K. Clark,
  Maksims Volkovs, and Leandro Aolita.
\newblock {Probabilistic simulation of quantum circuits using a deep-learning
  architecture}.
\newblock {\em Phys. Rev. A}, 104:032610, Sep 2021.

\bibitem{Werner2016positivetensor}
A.~H. Werner, D.~Jaschke, P.~Silvi, M.~Kliesch, T.~Calarco, J.~Eisert, and
  S.~Montangero.
\newblock {Positive Tensor Network Approach for Simulating Open Quantum
  Many-Body Systems}.
\newblock {\em Physical Review Letters}, 116:237201, Jun 2016.

\bibitem{Arute2019syc}
Frank Arute, Kunal Arya, Ryan Babbush, Dave Bacon, Joseph~C. Bardin, Rami
  Barends, Rupak Biswas, Sergio Boixo, Fernando G. S.~L. Brandao, David~A.
  Buell, Brian Burkett, Yu~Chen, Zijun Chen, Ben Chiaro, Roberto Collins,
  William Courtney, Andrew Dunsworth, Edward Farhi, Brooks Foxen, Austin
  Fowler, Craig Gidney, Marissa Giustina, Rob Graff, Keith Guerin, Steve
  Habegger, Matthew~P. Harrigan, Michael~J. Hartmann, Alan Ho, Markus Hoffmann,
  Trent Huang, Travis~S. Humble, Sergei~V. Isakov, Evan Jeffrey, Zhang Jiang,
  Dvir Kafri, Kostyantyn Kechedzhi, Julian Kelly, Paul~V. Klimov, Sergey Knysh,
  Alexander Korotkov, Fedor Kostritsa, David Landhuis, Mike Lindmark, Erik
  Lucero, Dmitry Lyakh, Salvatore Mandr{\`a}, Jarrod~R. McClean, Matthew
  McEwen, Anthony Megrant, Xiao Mi, Kristel Michielsen, Masoud Mohseni, Josh
  Mutus, Ofer Naaman, Matthew Neeley, Charles Neill, Murphy~Yuezhen Niu, Eric
  Ostby, Andre Petukhov, John~C. Platt, Chris Quintana, Eleanor~G. Rieffel,
  Pedram Roushan, Nicholas~C. Rubin, Daniel Sank, Kevin~J. Satzinger, Vadim
  Smelyanskiy, Kevin~J. Sung, Matthew~D. Trevithick, Amit Vainsencher, Benjamin
  Villalonga, Theodore White, Z.~Jamie Yao, Ping Yeh, Adam Zalcman, Hartmut
  Neven, and John~M. Martinis.
\newblock {Quantum supremacy using a programmable superconducting processor}.
\newblock {\em Nature}, 574(7779):505--510, Oct 2019.

\bibitem{Ho2019hva}
Wen~Wei Ho and Timothy~H. Hsieh.
\newblock {Efficient variational simulation of non-trivial quantum states}.
\newblock {\em SciPost Phys.}, 6:29, 2019.

\bibitem{Wierichs2020avoiding}
David Wierichs, Christian Gogolin, and Michael Kastoryano.
\newblock {Avoiding local minima in variational quantum eigensolvers with the
  natural gradient optimizer}.
\newblock {\em Phys. Rev. Research}, 2:043246, Nov 2020.

\bibitem{Wiersema2020exploring}
Roeland Wiersema, Cunlu Zhou, Yvette de~Sereville, Juan~Felipe Carrasquilla,
  Yong~Baek Kim, and Henry Yuen.
\newblock {Exploring Entanglement and Optimization within the Hamiltonian
  Variational Ansatz}.
\newblock {\em PRX Quantum}, 1:020319, Dec 2020.

\bibitem{Honeywell}
Honeywell Quantum.
\newblock {Mid-circuit measurements on the System Model H1}.
\newblock
  \url{https://www.honeywell.com/us/en/company/quantum/quantum-computer/},
  2021.
\newblock [Online; accessed 07/18/2021].

\bibitem{IBM}
IBM Quantum.
\newblock {Mid-circuit Measurements Tutorial}.
\newblock
  \url{https://quantum-computing.ibm.com/lab/docs/iql/manage/systems/midcircuit-measurement/},
  2021.
\newblock [Online; accessed 07/18/2021].

\bibitem{CHSH1969}
John~F. Clauser, Michael~A. Horne, Abner Shimony, and Richard~A. Holt.
\newblock {Proposed Experiment to Test Local Hidden-Variable Theories}.
\newblock {\em Phys. Rev. Lett.}, 23:880--884, Oct 1969.

\bibitem{Verstraete2004mpo}
F.~Verstraete, J.~J. Garc\'{\i}a-Ripoll, and J.~I. Cirac.
\newblock {Matrix Product Density Operators: Simulation of Finite-Temperature
  and Dissipative Systems}.
\newblock {\em Physical Review Letters}, 93:207204, Nov 2004.

\bibitem{Moore1920geninv}
Eliakim Moore.
\newblock {On the reciprocal of the general algebraic matrix}.
\newblock {\em Bulletin of the American Mathematical Society}, 26(9):394 --
  395, 1920.

\bibitem{Penrose1955geninv}
R.~Penrose.
\newblock {A generalized inverse for matrices}.
\newblock {\em Mathematical Proceedings of the Cambridge Philosophical
  Society}, 51(3):406--413, 1955.

\bibitem{Metropolis1953mh}
Nicholas Metropolis, Arianna~W. Rosenbluth, Marshall~N. Rosenbluth, Augusta~H.
  Teller, and Edward Teller.
\newblock {Equation of State Calculations by Fast Computing Machines}.
\newblock {\em The Journal of Chemical Physics}, 21(6):1087--1092, 1953.

\bibitem{Sherlock2010rwmh}
Chris Sherlock, Paul Fearnhead, and Gareth~O. Roberts.
\newblock {The Random Walk Metropolis: Linking Theory and Practice Through a
  Case Study}.
\newblock {\em Statistical Science}, 25(2):172 -- 190, 2010.

\bibitem{Gelman1997weakconv}
A.~Gelman, W.~R. Gilks, and G.~O. Roberts.
\newblock {Weak convergence and optimal scaling of random walk Metropolis
  algorithms}.
\newblock {\em The Annals of Applied Probability}, 7(1):110 -- 120, 1997.

\end{thebibliography}
  
  \begin{appendices}
    
  \section{Informationally Complete POVMs \label{app:icpovm}} 
  A positive operator valued measure (POVM) is a set of operators $\{M_{\boldsymbol{a}}\}_{\boldsymbol{a}}$ with $M_{\boldsymbol{a}}\geq 0$ that satisfies the condition
  \begin{align}
    \sum_{\boldsymbol{a}} M_{\boldsymbol{a}} = I.
  \end{align}
  A POVM is informationally complete if $\{M_{\boldsymbol{a}}\}_{\boldsymbol{a}}$ spans $\cL(\mathbb{H}_{\mathcal{S}})$. Let $\{{M}_{a_i}\}_{a_i}$ be a POVM that act on a single-qubit Hilbert space. We can define a factorable POVM as a tensor product of single-qubit POVM element as 
  \begin{align}
   {M}_{\boldsymbol{a}}={M}_{a_1}\otimes \hdots \otimes{M}_{a_N},
  \end{align}
  for $\boldsymbol{a}:=(a_1,\hdots a_N)$. Clearly, if all $M_{a_i}$ are informationally complete, then so is $M_{\boldsymbol{a}}$. An example of a informationally complete POVM is the Pauli-6 POVM, which is defined as
  \begin{align}
      \{M_{\boldsymbol{a}}\}_{\boldsymbol{a}}^{\text{Pauli-6}} := \bigcup_{a=x.y,z} \left\{\frac{1}{3}\ketbra{\uparrow_{a}}{\uparrow_{a}},\frac{1}{3}\ketbra{\downarrow_{a}}{\downarrow_{a}}\right\},
  \end{align}
  where the vectors $\ket{\uparrow_\alpha}, \ket{\downarrow_\alpha}$ correspond to the eigenvectors of the Pauli operators with eigenvalue $\pm 1$ respectively.

  \section{Dual frame decomposition \label{app:qpr}}
  Here, we show that Eq. \eqref{eq:cqi} defines a dual frame with respect to $\mathcal{F}_\mathcal{S}$ if Eq. \eqref{eq:t_tau_t} holds. For the forward direction of this statement, we start with Eq. \eqref{eq:frame_def} and plug in Eq. \eqref{eq:cqi} to obtain
  \begin{align}
   \mathcal{I}=\sum_{\boldsymbol{a},\boldsymbol{b}} \mathfrak{T}_{\boldsymbol{a},\boldsymbol{b}} \cket{M_{\boldsymbol{b}}} \cbra{M_{\boldsymbol{a}}}\label{eq:step_1}.
  \end{align}
  Applying $\cbra{M_{\boldsymbol{c}}}$ and $\cket{M_{\boldsymbol{d}}}$ to the left and right of Eq. \eqref{eq:step_1} then gives
  \begin{align}
   \cbraket{M_{\boldsymbol{c}}}{M_{\boldsymbol{d}}} = \sum_{\boldsymbol{a},\boldsymbol{b}} \mathfrak{T}_{\boldsymbol{a},\boldsymbol{b}} \cbraket{M_{\boldsymbol{c}}}{M_{\boldsymbol{b}}} \cbraket{M_{\boldsymbol{a}}}{M_{\boldsymbol{d}}}
  \end{align}
  therefore we see that $T = T\mathfrak{T}T$ as required.
  
  For the converse direction, we start with a map $\mathcal{J}$ on $\cL(\mathbb{H}_{\mathcal{S}})$,
  \begin{align}
   \mathcal{J} =\sum_{\boldsymbol{a},\boldsymbol{b}} \mathfrak{T}_{\boldsymbol{a},\boldsymbol{b}} \cket{M_{\boldsymbol{b}}} \cbra{M_{\boldsymbol{a}}}\label{eq:step_3}.
  \end{align}
  Applying $\cbra{M_{\boldsymbol{c}}}$ and $\cket{M_{\boldsymbol{d}}}$ to the left and right of Eq. \eqref{eq:step_3} then gives
  \begin{align}
   \cbra{M_{\boldsymbol{c}}}\mathcal{J}\cket{M_{\boldsymbol{d}}} =\sum_{\boldsymbol{a},\boldsymbol{b}} \mathfrak{T}_{\boldsymbol{a},\boldsymbol{b}} \cbraket{M_{\boldsymbol{c}}}{M_{\boldsymbol{b}}} \cbraket{M_{\boldsymbol{a}}}{M_{\boldsymbol{d}}}.
  \end{align}
  If we then plug in Eq. \eqref{eq:t_tau_t} we find
  \begin{align}
   \cbra{M_{\boldsymbol{c}}}\mathcal{J}\cket{M_{\boldsymbol{d}}} = \cbraket{M_{\boldsymbol{c}}}{M_{\boldsymbol{d}}},
  \end{align}
  from which we conclude that $\mathcal{J}\equiv\mathcal{I}$, i.e. $\mathcal{J}$ equals the identity map.

  \section{Finite statistics estimator \label{app:finite_stats}}
  Let $O$ be a generic observable we wish to measure, with support on an arbitrary subset of $\mathcal{S}$ and with arbitrary spectral norm $\|O\|_{\rm sp}\coloneqq o_{\rm max}$. Hence, it admits a spectral decomposition as $\cket{O}=\sum_{\lambda} o_{\lambda} \cket{\lambda}$, where $o_{\lambda}$ and $\cket{\lambda}$ are respectively its $\lambda$-th eigenvalue and eigenvector projector, with  $|o_{\lambda}|\leq o_{\rm max}$ for all $\lambda$.
  Using Eq. \eqref{eq:final_exp}, we write the finite statistics estimator of the expectation value $\langle O\rangle\coloneqq {\rm Tr}[O\, \varrho_{\rm f}]$ of $O$ as
  \begin{equation}
    \label{eq:estimateO}
    O^{*}_{M}\coloneqq \frac{1}{M}\sum_{i=1}^{M}o_{\lambda^{(i)},\boldsymbol{\alpha}_{s_L}^{(i)}}\prod_{k\in L} v_{\boldsymbol{a}_{s_k}^{(i)},\boldsymbol{b}_{s_k}^{(i)}},
  \end{equation}
  where $o_{\lambda^{(i)},\boldsymbol{\alpha}_{s_L}^{(i)}}$ is the eigenvalue obtained from the single-shot $i$ obtained from a state that is measured and reprepared according to $\boldsymbol{\alpha}_{s_L}^{(i)}$. The probability of observing $o_{\lambda^{(i)},\boldsymbol{\alpha}_{s_L}^{(i)}}$ is given by
  \begin{equation}
  P(o_{\lambda^{(i)},\boldsymbol{\alpha}_{s_L}^{(i)}}) = (\lambda^{(i)}| \prod_{k=1}^{f}  \mathcal{W}_k(\boldsymbol{a}^{(i)}_{s_k},\boldsymbol{b}^{(i)}_{s_k}) |\varrho_{0}),
  \end{equation}
  with $\boldsymbol{a}^{(i)}_{s_k}\sim P_{\varrho_{k-1}}(\boldsymbol{a}_{s_k})$ and $\boldsymbol{b}^{(i)}_{s_k}\sim P_{\mathcal{U}_k}(\boldsymbol{b}_{s_k}|\boldsymbol{a}_{s_k})$ where \begin{equation}
  \cket{\varrho_{k-1}} = \prod_{k=1}^{l-1}  \mathcal{W}_l(\boldsymbol{a}_{s_l}^{(i)},\boldsymbol{b}_{s_l}^{(i)}) |\varrho_{0}).
  \end{equation}
  Importantly, $O^{*}_{M}$ is an unbiased estimator. 

  \section{The Clauser-Horne-Shimony-Holt inequalities \label{app:chsh}}
  The CHSH inequalities constrain a set of four correlators in an Alice (A) and Bob (B) type experiment and provide a condition to check if the correlations between the observations of Alice and Bob can be explained by a local theory, or necessitate a non-local theory such as quantum mechanics \cite{CHSH1969}. Consider the quantity
  \begin{align}
    S(A,B) = &C^{00}(A,B) + C^{01}(A,B)+ C^{10}(A,B) \\
    &-C^{11}(A,B)
  \end{align}
  where 
  \begin{align}
    C^{00}(A,B) &= \frac{1}{\sqrt{2}}(-\expval{Z_{A} Z_B} - \expval{Z_{A} X_B})\\
    C^{01}(A,B) &= \frac{1}{\sqrt{2}}(-\expval{X_{A} Z_B} - \expval{X_{A} X_B})\\
    C^{10}(A,B) &= \frac{1}{\sqrt{2}}(\expval{Z_{A} Z_B} - \expval{Z_{A} X_B})\\
    C^{11}(A,B) &= \frac{1}{\sqrt{2}}(\expval{X_{A} Z_B} - \expval{Z_{A} X_B})
  \end{align}
  are the correlations obtained from the state shared by Alice and Bob. The observables $X$ and $Z$ are the Pauli matrices. We call $S(A,B)$ the Bell polynomial. The CHSH inequality is given by $S(A,B) \leq 2$, which if satisfied, implies that a local hidden variable theory can explain the observed correlations. On the other hand, for $S(A,B)>2$ we have to invoke quantum theory to explain the correlations. The maximum value of $S(A,B)$ is $2\sqrt{2}$ which is obtained for a maximally entangled two qubit state.
  
  \section{Locally Purified Density Operators\label{app:lpdo}}
  
  Numerical simulations with the full density matrix of size $2^N\times 2^N$ quickly become prohibitive due to the large memory requirements. Hence we have to resort to tensor networks to find efficient representations of mixed quantum states. The canonical choice for representing operators with tensor networks are matrix product operators (MPO) \cite{Verstraete2004mpo}. A drawback of this approach is that applying completely positive maps to the state can still lead to the MPO becoming non-positive due to truncation errors. The Locally Purified Density Operator tensor network solves this issue by representing the state as $\varrho = \chi\chi^\dag$, where the purification operator $\chi$ is given by a tensor network
  \begin{align}
    [\chi]^{p_1, \ldots, p_N}_{k_1, \ldots, k_N} = \sum_{b_1,\ldots,b_{N-1}} A^{[1] p_1, k_1}_{b_1} A^{[2] p_2, k_2}_{b_1,b_2}\ldots  A^{[N] p_N k_N}_{b_{N-1}},
  \end{align}
  with $1\leq p_l \leq P$, $1\leq k_l \leq K$ and $1\leq b_l \leq D$ \cite{Werner2016positivetensor}. Here, $P$ is called the physical dimension, $K$ is the Kraus dimension and $D$ is the bond dimension. 
  
\begin{figure}[htb!]
    \includegraphics[width=\columnwidth]{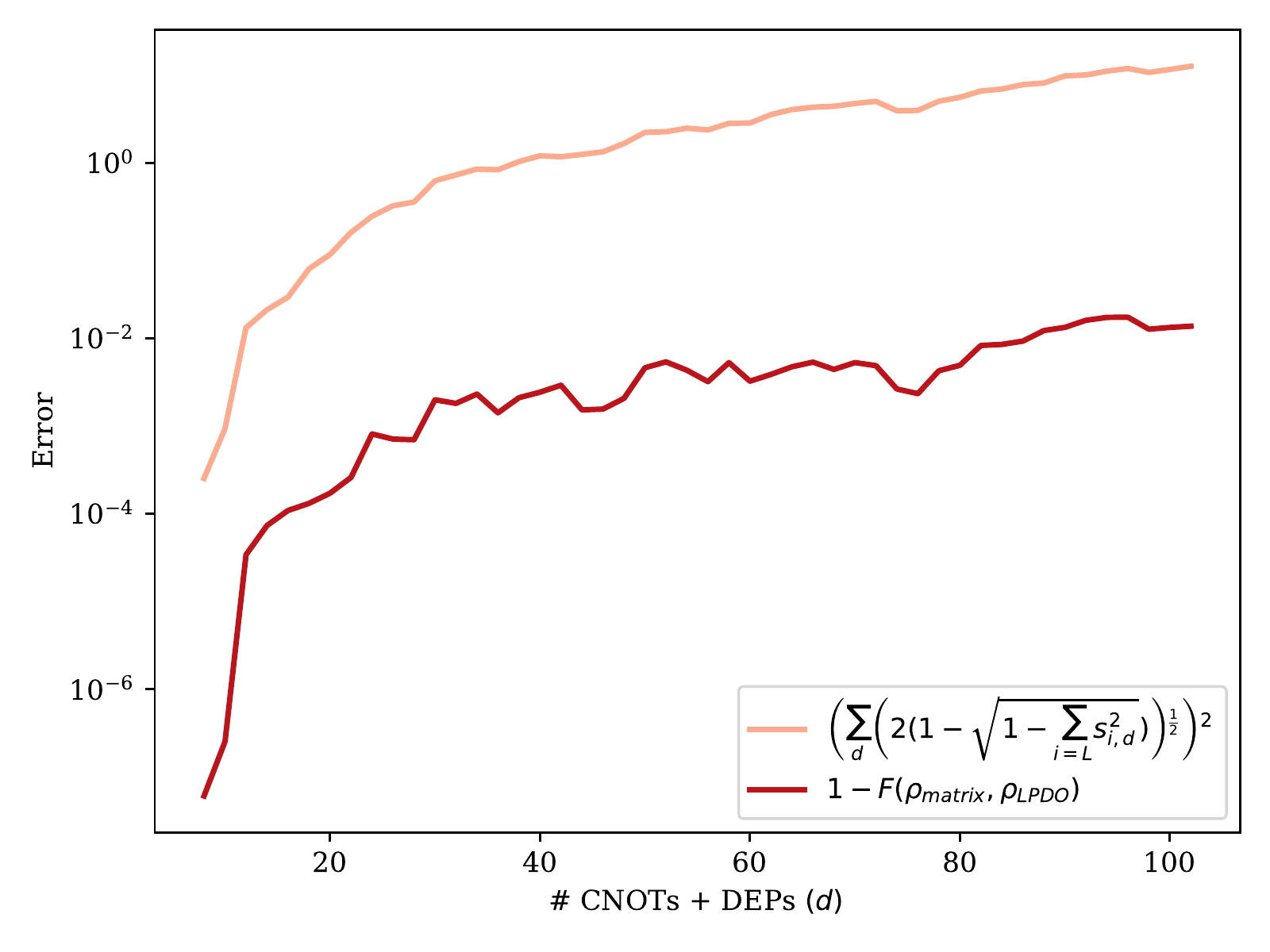}
    \caption{Illustration of the lower bound of Eq. \eqref{eq:bound}. The circuit consists of an initial state $\ket{+}^{\otimes4}$ to which we apply a varying number of CNOT gates with random control and target qubits. We set the noise to $\lambda=0.005$ and take $D=4$ and $K=16$. The red line indicates the true accuracy of the LPDO simulation by comparing with the exact full density matrix simulation. The orange line gives the runtime fidelity estimate. We see that the accuracy of the simulation degrades as we add more two-qubit gates and depolarizing channels. The runtime fidelity gives an estimate two orders of magnitude above the exact error, indicating that for this example, the bound is a conservative estimate of the simulation error. 
    }
    \label{fig:bound_estimate}
  \end{figure}

  Analogous to the bond dimension truncation in MPOs, truncating the Kraus dimension after applying a channel leads to errors in our state representation that can affect the accuracy of numerical simulations. However, we can control the accuracy of the simulation by increasing $D$ and $K$ and keeping track of a runtime lower bound estimate of the state fidelity. Let $\varrho = \chi^\dag \chi$, $\sigma=\eta^\dag \eta$, then the fidelity is given by
  \begin{align}
      F(\varrho, \sigma) = \Tr{\sqrt{\sqrt{\sigma}\varrho\sqrt{\sigma}}}.
  \end{align}
  From Lemma 1 in \cite{Werner2016positivetensor} we know that, 
  \begin{align}
      F(\varrho, \sigma) \geq \frac{1}{2}\left(2 - \norm{\chi - \eta}_2^2\right).
  \end{align}
  Let $\chi$ be a locally purified description of a quantum state with local tensors $\{A^{[N]}\}$ that is in mixed canonical form with respect to a local tensor $A^{[l_{cp}]}$. If a single tensor $A^{[l]}$ is compressed by discarding singular values in either the Kraus or bond dimensions, then by Lemma 6 of \cite{Werner2016positivetensor} we know that
  \begin{align}
      \delta := \left(\sum_{i, \text{discarded}} s_i^2 \right)^{\frac{1}{2}},
  \end{align}
  and subsequently
  \begin{align}
      \norm{\chi - \chi'}_2^2 = 2(1 - \sqrt{1 - \delta^2})
  \end{align}
  where $\chi'$ is the compressed tensor.
  By the triangle inequality, the two norm errors introduced by the discarded weights can at most sum up. Hence the true operator norm is lower bounded by the sum of all discarded weight errors
  \begin{align}
      \norm{\varrho_{\text{exact}} - \varrho_{\text{truncated}}}_2 \leq \sum_{d}  \sqrt{2(1 - \sqrt{1 - \delta_d^2})}
  \end{align}
  With $d$ the number of truncations and $\delta_k$ the discarded weights. This brings the final runtime fidelity estimate to
  \begin{align}
      F(\varrho, \sigma) & \geq \frac{1}{2}\left(2 - \norm{\chi - \eta}_2^2\right) \\
      &\geq \frac{1}{2}\left(2 - \left(\sum_{d}  \sqrt{2(1 - \sqrt{1 - \delta_d^2})}\right)^2\right)\label{eq:bound}
  \end{align}
  In all our experiments, we apply depolarizing channels to both qubits only after applying a two qubit gate, since single qubit gate noise tends to be small in experimental settings. The single-qubit depolarizing channel is given by
  \begin{align}
    \mathcal{\varrho} = \sum_{m=1}^M K_m \varrho K_m^\dag,
  \end{align}
  where $\{K_m\}$ is a set of Kraus operators with
  \begin{align}
    K_1 &= \sqrt{\frac{(4 - 3 \lambda)}{4}}  \openone,\quad
    K_2 = \sqrt{\frac{\lambda}{ 4}} X\\
    K_3 &= \sqrt{\frac{\lambda}{ 4}} Y, \quad
    K_4 =  \sqrt{\frac{\lambda}{ 4}} Z.
  \end{align}
  Here, $\{X,Y,Z\}$ are the Pauli matrices and $\openone$ is the identity.
  The scalar $\lambda\in[0,1]$ controls the strength of the depolarization. 
  With these channels, illustrate the bound of Eq. \eqref{eq:bound} by comparing the final state overlap of an exact full density matrix simulation and a LPDO simulation for a random 4 qubit circuit with a varying number of CNOT gates. In Fig. \ref{fig:bound_estimate}, we see that the runtime estimate of the fidelity is about two orders of magnitude above the true fidelity.
  
  \section{Random Walk Metropolis-Hastings for negativity minimization \label{app:neg_min} }
 In this appendix, we present a method to minimize the sample-complexity overhead by the interface of a unitary gate $U$ exploiting the freedom in the choice of dual frame, namely the choice of $\mathfrak{T}$ subject to Eq. \eqref{eq:t_tau_t}. For concreteness, we focus on the case where all POVM elements have the same trace, so $t_{\boldsymbol{b}}=1/D$ for all $\boldsymbol{b}$, with $D$ the number of POVM elements. Moreover, we optimize a modified version of the interface negativity $n_{U}$ where, instead of maximizing $\big\|\tilde{T}^{(\mathcal{U})}_{\boldsymbol{a}}\big\|_1$ over $\boldsymbol{a}$ [as in Eq. \eqref{eq:def_neg}], we average $\big\|\tilde{T}^{(\mathcal{U})}_{\boldsymbol{a}}\big\|^2_1$ over $\boldsymbol{a}$. Such an average is the sample-complexity overhead directly given by the Hoeffding bound for when the sampled random variables can lie within segments of different lengths. The reason for this modification is that, while in Theo. \ref{theorem1} we are interested in the worst-case complexity, here we are interested in the more practical problem of the average case.
 
For optimizing $\tilde{T}^{(\mathcal{U})}$ over $\mathfrak{T}$, we express it as $\tilde{T}^{(\mathcal{U})} = \mathfrak{T}_1\, T^{(\mathcal{U})}\mathfrak{T}_2$, with $T^{(\mathcal{U})}$ given by $T^{(\mathcal{U})}_{\boldsymbol{b},\boldsymbol{a}}:= (M_{\boldsymbol{b}}|\,\mathcal{U}\,|M_{\boldsymbol{a}})$. Note that by not enforcing that $\mathfrak{T}_1=\mathfrak{T}_2$, we are explicitly allowing for the more general case of possibly different input and output dual frames.
Hence, we wish to solve the constrained non-convex optimization 
  \begin{align}
    \min_{\mathfrak{T}}  \quad &\frac{1}{D}\sum_{\boldsymbol{a}}\big\|\big(\mathfrak{T}_1\, T^{(\mathcal{U})}\mathfrak{T}_2
    \big)_{\boldsymbol{a}}\big\|^2_1,
    \label{eq:optimization_neg}\\
    &\text{s.t.}\quad T = T \mathfrak{T}_i T, \text{ for } i=1,2,
  \label{eq:opt_constraint}
  \end{align}
where $\big(\mathfrak{T}_1\, T^{(\mathcal{U})}\mathfrak{T}_2 \big)_{\boldsymbol{a}}$ is a short-hand notation for the $\boldsymbol{a}$-th row of $\mathfrak{T}_1\, T^{(\mathcal{U})}\mathfrak{T}_2$ and $\big\|\big(\mathfrak{T}_1\, T^{(\mathcal{U})}\mathfrak{T}_2
    \big)_{\boldsymbol{a}}\big\|_1$ its $l_1$-norm.
Eq. \eqref{eq:opt_constraint} is a necessary but not sufficient condition for $\mathfrak{T}_i$ to be the Penrose-Moore pseudo-inverse of $T$. Indeed, such condition implies that $\mathfrak{T}_i$ is a so-called generalized inverse of $T$ \cite{Moore1920geninv, Penrose1955geninv}. So, the first question we need to consider is how to variationally explore the space of generalized inverses of $T$ in a practical way.   
  
  
Fortunately, this question has been previously studied. In particular, in Ref. \cite{Rao1967geninv} it was shown that for an arbitrary matrix $A\in\mathbb{R}^{m\times n}$ and given any particular generalized inverse $A^{-}$ of it, every generalized inverse $B^{-}$ can be obtained from some $C\in\mathbb{R}^{m\times n}$ by the map 
  \begin{align}
  B^{-}(C):= A^{-} + C - A^{-}ACAA^{-}
  \label{eq:generate_gis}.
  \end{align}
That is, the entire space of generalized inverses is parametrized by $C$. This leads us to a practical way to obtain a random walk across the space of generalized inverses: In the first iteration, take the Penrose-Moore pseudo-inverse $A^{-1}$ as starting generalized inverse  and a randomly sampled $C$. This produces the first $B^{-}$. As inputs for the second iteration, use the firsts iteration's output $B^{-}$ as generalized inverse and a fresh, independently sampled C. This produces a new $B^{-}$. Then continue to iterate.
  
Using this recipe for $A = T$ and $A^{-1}=T^{-1}$, we can ergodically explore the space of generalized inverses $\mathfrak{T}_i$ of $T$. In turn, the resulting random walk can be used as Markov Chain Monte Carlo dynamics for a simulated-annealing optimization  \cite{Metropolis1953mh, Sherlock2010rwmh} that approximates a solution to Eq. \eqref{eq:optimization_neg}. More precisely, for each random walk iteration, we (probabilistically)  accept or reject the newly produced $\mathfrak{T}_i$ via a standard Metropolis-Hastings algorithm with $\frac{1}{D}\sum_{\boldsymbol{a}}\big\|\big(\mathfrak{T}_1\, T^{(\mathcal{U})}\mathfrak{T}_2\big)_{\boldsymbol{a}}\big\|^2_1$ as energy function.
  
For $U$ a two-qubit gate and the Pauli-6 POVM as primal frame, each dual-overlap matrix can be expressed as $\mathfrak{T}_i=\mathfrak{T}^{(1)}_i\otimes\mathfrak{T}^{(2)}_i$, where $\mathfrak{T}^{(1)}_i$ and $\mathfrak{T}^{(2)}_i$ are respectively the $6\times 6$ real dual-overlap matrices of the two qubits on which $U$ acts. We can independently sample all four matrices, $\mathfrak{T}^{(1)}_1$, $\mathfrak{T}^{(2)}_1$, $\mathfrak{T}^{(1)}_2$, and $\mathfrak{T}^{(2)}_2$. Hence the search-space dimension is $4\times 6\times6=142$. 
 For the simulated-annealing schedule, we take random matrices $C \sim \mathcal{N}(0,\sigma^2)^{6\times 6}$. We set the initial temperature to be $T=10$ and decrease it with a factor $0.999$ at each Monte Carlo step. In addition to the temperature, the Monte dynamics are controlled by the variance $\sigma^2$ of the normal distribution $\mathcal{N}(0,\sigma^2)^{6\times 6}$ for $C$. We start with a large initial $\sigma^2=0.1$ to coarsely explore the search space. However, as the temperature decreases, we want to refine the search without freezing the Monte Carlo dynamics. Therefore, we use an adaptive scheme where $\sigma^2$ is decreased according to the acceptance ratio. Specifically, we halve the value of $\sigma$ if the acceptance ratio per 100 MCS is smaller than $0.23$, a well-known heuristic for continuous-variable MCMC \cite{Gelman1997weakconv}. The search is terminated if the negativity decreases less than $10^{-2}$ after 100 accepted steps.
  
\begin{figure*}[htb!]
    \includegraphics[width=\textwidth]{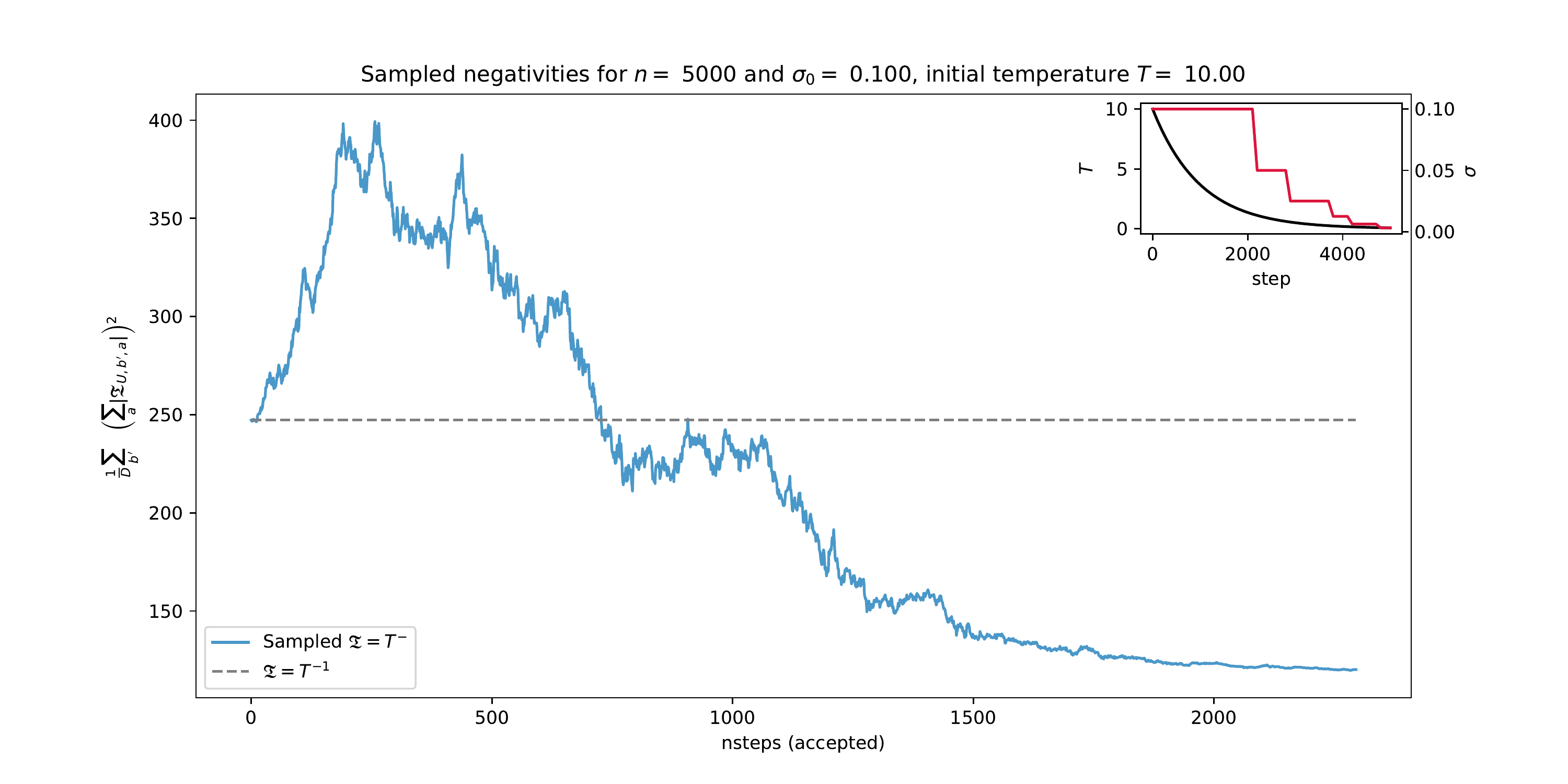}
    \caption{Monte Carlo random walk for interface negativity optimization of the $ZZ$ gate used in Sec. \ref{sec:TFIsing}. The gray dashed line indicates the mean average squared negativity of the pseudo-inverse, whereas the blue line indicates the one for the newly accepted $\mathfrak{T}$'s during the Monte Carlo random walk. The inset shows the adaptive scheme that fine-tunes the search with the temperature and variance, given in black and red, respectively.
    }
    \label{fig:mcmcrw}
  \end{figure*}

\begin{figure*}[htb!]
    \includegraphics[width=0.7\textwidth]{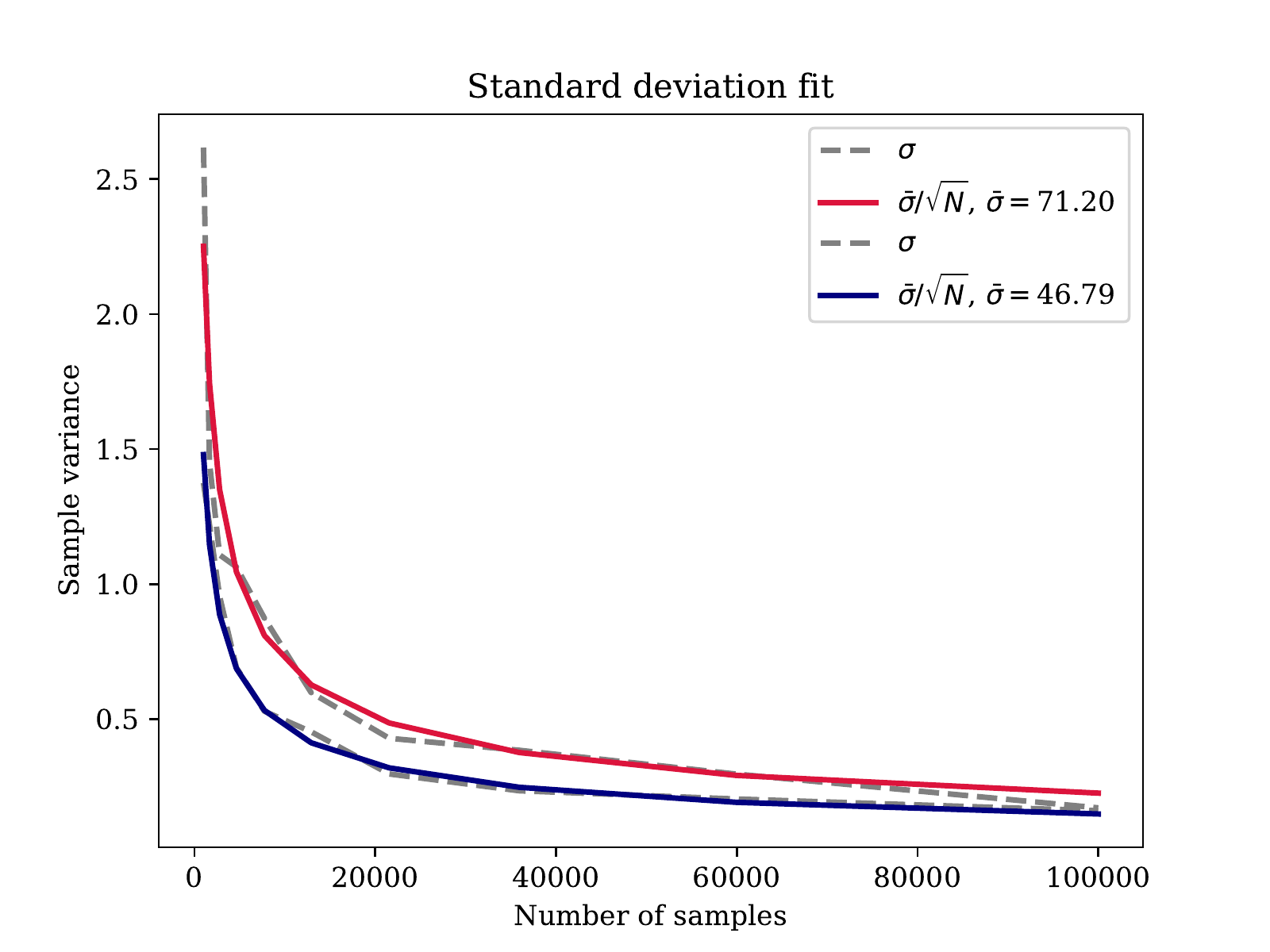}
    \caption{Improvement of energy-estimator variance for the 8-qubit TFIM circuit experiment of Fig. \ref{fig:tfim_8_single}. Sample variance is estimated over 50 runs. The red line shows sample variance corresponding to the canonical dual frame given by the pseudo-inverse of $T$. In blue we see the variance of the energy corresponding to the dual frame obtained from the Monte Carlo search. 
    }
    \label{fig:sample_var}
  \end{figure*}

As a result, we consistently find dual frames whose averaged squared negativites are about half the value of the canonical dual frame from the pseudo-inverse (see Fig. \ref{fig:mcmcrw}). This is also observed to greatly improve the sample complexity in practice (see Fig. \ref{fig:sample_var}).
%
  
  \end{appendices}
\end{document}